\documentclass{ieeeaccess}

\usepackage{cite}
\usepackage{amsmath,amssymb,amsfonts}
\usepackage{algorithm}
\usepackage{algpseudocode}   
\usepackage{graphicx}
\usepackage{textcomp}
\usepackage{booktabs}
\usepackage{array}
\newcolumntype{L}[1]{>{\raggedright\arraybackslash}p{#1}}
\usepackage{tabularray}      
\usepackage{subcaption}
\usepackage{xcolor}
\usepackage{url}
\usepackage{threeparttable}   
\usepackage{nicefrac}
\usepackage{microtype}

\renewcommand{\dbltopfraction}{0.9}
\renewcommand{\dblfloatpagefraction}{0.8}
\renewcommand{\topfraction}{0.9}
\newcommand{\MAES}{\mathrm{MAE}_S}

\DeclareMathOperator{\Tr}{Tr}
\newcommand{\ket}[1]{\left| #1 \right\rangle}

\makeatletter
\def\headerlogo{}
\def\headerlogoall{}
\def\ps@preprint{%
  \def\@oddhead{\hbox to \textwidth{\headerfont\rightmark\hfil}}%
  \let\@evenhead\@oddhead
  \def\@oddfoot{\hfil{\footerpagefont\thepage}\hfil}%
  \let\@evenfoot\@oddfoot
}
\def\ps@titlepage{%
  \let\@oddhead\@empty
  \let\@evenhead\@empty
  \def\@oddfoot{\hfil{\footerpagefont\thepage}\hfil}%
  \let\@evenfoot\@oddfoot
}
\def\EOD{\global\eoddefinedtrue}
\AtBeginDocument{\pagestyle{preprint}}
\makeatother

\begin{document}

\history{}
\doi{}

\title{Energy Accuracy Is Not Enough: A Structure-Aware Benchmark and
Evaluation Protocol for Quantum Architecture Search}

\author{
\uppercase{Jiayang Niu}\authorrefmark{1,*},
\uppercase{Akib Karim}\authorrefmark{2,*},
\uppercase{Yan Wang}\authorrefmark{1},
\uppercase{Jie Li}\authorrefmark{1},
\uppercase{Ke Deng}\authorrefmark{1},
\uppercase{Azadeh Alavi}\authorrefmark{1},
\uppercase{Muhammad Usman}\authorrefmark{3},
\uppercase{and Yongli Ren}\authorrefmark{1}}

\address[1]{School of Computing Technologies, RMIT University, Melbourne,
VIC 3000, Australia (e-mail: s4068570@student.rmit.edu.au;
yongli.ren@rmit.edu.au)}
\address[2]{Quantum Systems, Data61, CSIRO, Australia
(e-mail: akib.karim@csiro.au)}
\address[3]{Faculty of Information Technology, Monash University, Clayton,
VIC 3800, Australia (e-mail: muhammad.usman@monash.edu)}

\tfootnote{This research is supported by the ARC Discovery Project (DP210100743) and RACE (RMIT Advanced Computing Ecosystem).}

\markboth
{Niu \headeretal: Energy Accuracy Is Not Enough}
{Niu \headeretal: Energy Accuracy Is Not Enough}

\corresp{Corresponding author: Yongli Ren (e-mail: yongli.ren@rmit.edu.au).\\
$^{*}$Jiayang Niu and Akib Karim contributed equally to this work.}

\begin{abstract}
Quantum architecture search for molecular ground-state estimation is commonly evaluated through energy accuracy, which does not describe circuit cost or the physical properties of the prepared state. We introduce HamQASBench, a structure-aware benchmark comprising eleven molecular Hamiltonians of up to fourteen qubits, selected using Hamiltonian and target-state properties and supplied with exact references. Its evaluation protocol combines energy accuracy and success rates with reference-relative circuit cost, local entropy profiles for non-degenerate targets, and state identification within degenerate ground subspaces. Experiments with five methods spanning four search paradigms reveal differences hidden by energy-only comparisons. On a near-product instance under the shallow search budget, the best outputs of all five methods reach chemical accuracy while using between two and sixty-two gates. Equal-energy outputs on a degenerate instance occupy distinct spin components. Local entropy profiles distinguish inaccurate outputs and show that entangling-gate counts need not reflect realized entanglement. Across the molecular instance ladder, product-state outputs can meet or miss chemical accuracy, motivating interpretation of success alongside target-state properties. These results support retaining energy as the task-success criterion while using circuit-cost and state diagnostics for more informative comparisons. The benchmark instances, references, evaluation implementation, and per-run data are released for reuse.

\end{abstract}


\titlepgskip=-15pt

\maketitle

{
\section{Introduction}
\label{sec:introduction}
\PARstart{Q}{uantum} architecture search (QAS) automates the design of parameterized circuits for quantum algorithms~\cite{zhang2022differentiable, patelcurriculum, niu2025hybrid, nakaji2024generative, wu2023quantumdarts}. For molecular ground-state problems, it searches for a circuit and parameters that approximate the ground-state energy of a given Hamiltonian. Energy error measures progress towards this goal, and chemical accuracy ($1.6\,\mathrm{mHa}$) provides a common success threshold. These measures leave two questions open: how much circuit cost is needed to reach the threshold, and what physical features does the prepared state have?

Circuit cost remains important when many methods reach chemical accuracy. On \texttt{BeH2\_STO3G}, the best outputs of all five methods in our shallow-budget experiments reach this threshold, yet their circuits range from two to sixty-two gates. A verified two-gate reference shows that a compact circuit meets the same threshold and provides a basis for comparing the costs of successful outputs. We also need to understand what state a circuit prepares. When several ground states share the same energy, an accurate energy estimate does not tell us which one was obtained. For example, two outputs may have the same ground-state energy but opposite spin projections. Comparing them with the known ground subspace and measuring their spin can distinguish these outcomes.

When the ground state is unique, we can instead examine how the output differs from that target. Single-qubit entropy measures how strongly each qubit is entangled with the rest of a pure state. Comparing these entropies with the exact target values shows which qubits have too little or too much entanglement. Counting CNOT gates cannot provide this information: their effect depends on the circuit parameters, and even a circuit containing many CNOTs can produce an unentangled state. These comparisons require known properties of the target, such as ground-state degeneracy and local entropies. We therefore use these properties to select benchmark instances that help us interpret the states returned by a search.

Existing QAS benchmarks support comparisons across tasks and search spaces~\cite{lu2023qas, ikhtiarudin2025benchrl, martyniuk2025benchmarking}. Hamiltonian interactions and target-state entanglement have also been used to guide ansatz design~\cite{wiersema2020exploring, joch2025entanglement}. We use Hamiltonian and target-state information to select benchmark instances and interpret search outputs. Our benchmark, \textsc{HamQASBench}, evaluates the energy accuracy, circuit cost, and prepared states of the returned circuits.

The article makes three contributions.

\emph{1) A diagnostic suite (Section~\ref{sec:suite}).} We provide eleven molecular Hamiltonians of up to 14 qubits, with exact spectral and ground-state references. Five diagnostic regimes examine circuit cost, state selection under degeneracy, local entropy differences, connectivity constraints, and scaling within a molecular family. Each regime states its question and comparison conditions.

\emph{2) An evaluation protocol (Section~\ref{sec:protocol}).} We combine energy error and success rate with three additional measurements: gate count relative to a compact reference circuit,
local entropy differences from a unique ground-state target, and identification of outputs within a degenerate ground subspace. The protocol specifies which outputs are evaluated and how results
are aggregated across runs.

\emph{3) An empirical study (Section~\ref{sec:baselines}).} We evaluate five methods spanning four search paradigms. The results show that chemically accurate circuits can differ substantially in cost and can prepare different states at the same ground-state energy. Local entropy profiles reveal how outputs differ from their targets under different connectivity constraints. Along the scaling ladder, product-state outputs meet chemical accuracy on some instances but miss it on others.
}
{
\section{Background and Related Work}
\label{sec:background}
\subsection{Quantum Architecture Search for Molecular Ground States}
\label{sec:bg_qas}
\label{sec:bg_vqe}

For a molecular Hamiltonian $H$ mapped to $n$ qubits, the ground-state problem seeks a state with energy close to the lowest eigenvalue $E_0$. A variational quantum eigensolver (VQE) prepares a state $|C(\theta)\rangle$ using a circuit structure $C$ and parameters $\theta$. A classical optimizer adjusts the parameters to minimize
\begin{equation}
E(C,\theta)=\langle C(\theta)|H|C(\theta)\rangle \ge E_0 .
\label{eq:vqe}
\end{equation}
In a conventional VQE calculation, the circuit structure is chosen in advance. Quantum architecture search (QAS) searches for this structure as well as its parameters. The methods evaluated here
represent four approaches to finding low-energy circuits.

\emph{Reinforcement-learning methods} build circuits through sequential decisions. An agent chooses which gate to add and learns from energy-based rewards~\cite{ostaszewski2021reinforcement, patelcurriculum, kundu2025tensorrl, niu2025hybrid}.

\emph{Differentiable methods} replace discrete gate choices with continuous architecture weights. They optimize these weights together with circuit parameters, then select a discrete circuit~\cite{zhang2022differentiable, wu2023quantumdarts}.

\emph{Candidate sampling and filtering methods} generate candidate circuits and rank them using inexpensive scores. Parameter optimization is then applied to a selected subset. For example, training-free QAS uses proxy scores to filter candidates before evaluating their energies~\cite{he2024training}.

\emph{Generative methods} learn a distribution over circuits and sample candidates from it. The model is trained to favor circuits suited to the target task~\cite{nakaji2024generative, furrutter2024quantum}.

Other QAS approaches use weight sharing to reduce candidate-evaluation cost~\cite{tang2021qubit} or tree search to explore circuit choices~\cite{wang2023automated}. Circuit representations also vary: methods may search over individual gates, blocks, or larger gadgets~\cite{olle2025scaling, kundu2024reinforcement}. Section~\ref{sec:procedure} describes the five implementations evaluated here, their search spaces, and their resource constraints.

\subsection{Benchmarks for Quantum Architecture Search}
\label{sec:bg_bench}

Comparing these search methods requires common test problems and evaluation criteria. Existing QAS benchmarks provide these in different settings. Lu et al.~\cite{lu2023qas} introduced a benchmark based on randomly generated unitary targets. BenchRL-QAS~\cite{ikhtiarudin2025benchrl} compared nine reinforcement-learning algorithms across several variational tasks. SQuASH~\cite{martyniuk2025benchmarking} used surrogate models to accelerate evaluation over predefined search spaces. Tabular and surrogate benchmarks are also widely used in neural architecture
search~\cite{ying2019bench, dong2020bench, klyuchnikov2022bench, mehta2022bench}.

Our benchmark focuses on interpreting the circuits returned on molecular ground-state tasks. We select instances with known target properties so that their outputs can be examined against suitable
references. For example, a known degenerate ground subspace lets us distinguish states with the same energy, while an exact local entropy profile shows how an output differs from a unique target.
Hamiltonian and target-state information therefore guides both instance selection and output analysis.

\subsection{Hamiltonian and Target-State Information}
\label{sec:bg_structure}

The information needed for these comparisons has also been used to guide circuit design. At the operator level, the Hamiltonian Variational Ansatz constructs circuits from the Hamiltonian's
interaction terms~\cite{wiersema2020exploring}. The structure of the problem thus helps determine which operations appear in the circuit.

At the state level, previous studies examine how target-state entanglement relates to circuit design, trainability, and VQE convergence~\cite{woitzik2020entanglement, leone2024practical}.
Per-qubit von Neumann entropy has also been used to guide ansatz construction~\cite{joch2025entanglement}. It describes how entanglement is distributed across the qubits of a pure target state.

These studies motivate using Hamiltonian and target-state properties to characterize benchmark instances. In \textsc{HamQASBench}, operator descriptors summarize the Hamiltonian, while exact spectra
and ground states establish references for analyzing search outputs. Circuit cost is assessed using compact circuits whose energies have been verified. The next section explains how we obtain these
references and select instances for the five diagnostic regimes.
}
{
\section{Hamiltonian-Informed Benchmark Construction}
\label{sec:suite}
\label{sec:construction}

We use Hamiltonian and target-state properties to organize eleven molecular instances into five diagnostic tasks. We first introduce the quantities used to describe these properties, then explain how we construct the instances and obtain their references.

\subsection{Hamiltonian and Target-State Descriptors}
\label{sec:fingerprints}

We characterize each instance by its Hamiltonian structure, ground-state spectrum, and target-state entanglement. These describe the operator's composition, whether the ground state is unique or degenerate, and how strongly each qubit is entangled with the rest of the target state. Table~\ref{tab:descriptors} summarizes the quantities used.

\paragraph{Hamiltonian structure.}
We describe the Hamiltonian through its Pauli terms and its matrix entries in the computational basis. In the Pauli expansion $H=\sum_i c_iP_i$, each $P_i$ is a product of single-qubit Pauli operators. The ratio $r_Z$ measures the fraction of absolute coefficient weight carried by diagonal terms, which contain only $I$ and $Z$. The ratio $r_{\ge2}$ measures the fraction carried by terms that act nontrivially on at least two qubits. Both ratios exclude the identity term and describe the composition of the Hamiltonian.

The matrix descriptors compare each diagonal entry with the sum of the magnitudes of the off-diagonal entries in its row. $G_1$ is the fraction of rows in which the diagonal magnitude is at least this sum. $G_2$ is the smallest ratio of these magnitudes over rows with a nonzero off-diagonal sum. We compute them after subtracting the mean diagonal value.

\paragraph{Ground-state spectrum.}
The spectrum determines whether the target is a unique ground state or a degenerate ground subspace. We order the eigenvalues as $E_0\le E_1\le\cdots$, retaining repeated values, and define $\mathrm{Gap}=E_1-E_0$. Under this definition, the gap is zero when the ground state is degenerate. For these instances, we also report the ground-state degeneracy, such as two-fold or three-fold, to specify the dimension of the reference subspace.

\paragraph{Target-state entanglement.}
For an exact reference state $|\psi_0\rangle$, the entropy of qubit $q$ is
\begin{equation}
S(q)=-\Tr(\rho_q\log_2\rho_q),\qquad
\rho_q=\Tr_{\bar q}(|\psi_0\rangle\langle\psi_0|).
\label{eq:vn_entropy}
\end{equation}
Here $\rho_q$ is the reduced density matrix obtained by tracing out all other qubits. For a pure state, $S(q)$ measures the entanglement of that qubit with the rest: zero means it is unentangled, and the maximum is one bit~\cite{nielsen2010quantum}. We record the full profile $(S(q_0),\ldots,S(q_{n-1}))$, its mean $\bar S_{\mathrm{exact}}$, and its maximum $S_{\max}=\max_k S(q_k)$. The profile shows differences between targets that can be hidden by similar mean entropies. All profiles use the instance's orbital basis and qubit mapping; for a degenerate target, we also specify the reference state.

\begin{table}[t]
\centering
\footnotesize
\caption{Three aspects of instance characterization.}
\label{tab:descriptors}
\setlength{\tabcolsep}{3pt}
\renewcommand{\arraystretch}{1.15}
\begin{tabular}{@{}L{0.25\columnwidth}L{0.27\columnwidth}L{0.42\columnwidth}@{}}
\toprule
Aspect & Quantities & Information provided \\
\midrule
Hamiltonian structure & $r_Z$, $r_{\ge2}$; $G_1$, $G_2$ & Pauli-term composition and relative diagonal/off-diagonal magnitudes \\
\addlinespace[4pt]
Ground-state spectrum & Gap; ground-state degeneracy & Unique or degenerate target and ground-subspace dimension \\
\addlinespace[4pt]
Target-state entanglement & $S(q)$, $\bar S_{\mathrm{exact}}$, $S_{\max}$ & Per-qubit entanglement, its mean, and its maximum \\
\bottomrule
\end{tabular}
\end{table}

\begin{table*}[t]
\centering
\footnotesize
\caption{Benchmark instances and exact target-state properties. $n$ is the qubit count; Deg.\ is the ground-subspace dimension. All entropies are in bits. Profiles are rounded to three decimal places; means are computed from the displayed values. A dash indicates an absent qubit. \texttt{H4\_Chain} serves both R3 and R4 and is listed once in each panel. Task labels R1--R5 are defined in Section~\ref{sec:regimes}.}
\label{tab:criteria_audit}
\setlength{\tabcolsep}{5pt}
\renewcommand{\arraystretch}{1.12}
\textbf{(a) Instance properties}\\[4pt]
\begin{tabular}{@{}llrrrr@{}}
\toprule
Task & Instance & $n$ & Gap (mHa) & Deg. & $\bar S_{\mathrm{exact}}$ \\
\midrule
R1 & \path{BeH2_STO3G} & 6 & 212 & 1 & 0.006 \\
R1 & \path{LiH_Equil} & 6 & 77 & 1 & 0.012 \\
R2 & \path{CH2}$^{a}$ & 8 & 0 & 3 & 0.017 \\
R3 & \path{H2_Stretch} & 4 & 4 & 1 & 0.974 \\
R3 & \path{H2O_StrongCorr} & 8 & 94 & 1 & 0.188 \\
R3, R4 & \path{H4_Chain} & 8 & 233 & 1 & 0.198 \\
R4 & \path{H3_Linear}$^{b}$ & 6 & 0 & 2 & 0.161 \\
R5 & \path{BeH2_631G} & 8 & 90 & 1 & 0.012 \\
R5 & \path{BeH2_6311G} & 10 & 63 & 1 & 0.005 \\
R5 & \path{BeH2_CCPVDZ_12q} & 12 & 68 & 1 & 0.008 \\
R5 & \path{BeH2_CCPVDZ_14q} & 14 & 61 & 1 & 0.019 \\
\bottomrule
\end{tabular}

\medskip
\textbf{(b) Single-qubit entropy profiles}\\[4pt]
\setlength{\tabcolsep}{2.5pt}
\begin{tabular}{@{}l*{14}{r}@{}}
\toprule
Instance & $q_{0}$ & $q_{1}$ & $q_{2}$ & $q_{3}$ & $q_{4}$ & $q_{5}$ & $q_{6}$ & $q_{7}$ & $q_{8}$ & $q_{9}$ & $q_{10}$ & $q_{11}$ & $q_{12}$ & $q_{13}$ \\
\midrule
\path{BeH2_STO3G} & 0.008 & 0.008 & 0.005 & 0.005 & 0.005 & 0.005 & -- & -- & -- & -- & -- & -- & -- & -- \\
\path{LiH_Equil} & 0.018 & 0.018 & 0.006 & 0.006 & 0.013 & 0.013 & -- & -- & -- & -- & -- & -- & -- & -- \\
\path{CH2}$^{a}$ & 0.000 & 0.000 & 0.000 & 0.068 & 0.000 & 0.000 & 0.000 & 0.068 & -- & -- & -- & -- & -- & -- \\
\path{H2_Stretch} & 0.974 & 0.974 & 0.974 & 0.974 & -- & -- & -- & -- & -- & -- & -- & -- & -- & -- \\
\path{H2O_StrongCorr} & 0.021 & 0.021 & 0.342 & 0.342 & 0.316 & 0.316 & 0.071 & 0.071 & -- & -- & -- & -- & -- & -- \\
\path{H4_Chain} & 0.124 & 0.124 & 0.274 & 0.274 & 0.284 & 0.284 & 0.109 & 0.109 & -- & -- & -- & -- & -- & -- \\
\path{H3_Linear}$^{b}$ & 0.202 & 0.139 & 0.147 & 0.147 & 0.083 & 0.245 & -- & -- & -- & -- & -- & -- & -- & -- \\
\path{BeH2_631G} & 0.024 & 0.024 & 0.002 & 0.002 & 0.002 & 0.002 & 0.021 & 0.021 & -- & -- & -- & -- & -- & -- \\
\path{BeH2_6311G} & 0.011 & 0.011 & 0.001 & 0.001 & 0.001 & 0.001 & 0.008 & 0.008 & 0.002 & 0.002 & -- & -- & -- & -- \\
\path{BeH2_CCPVDZ_12q} & 0.022 & 0.022 & 0.002 & 0.002 & 0.002 & 0.002 & 0.012 & 0.012 & 0.003 & 0.003 & 0.008 & 0.008 & -- & -- \\
\path{BeH2_CCPVDZ_14q} & 0.039 & 0.039 & 0.023 & 0.023 & 0.014 & 0.014 & 0.014 & 0.014 & 0.018 & 0.018 & 0.009 & 0.009 & 0.016 & 0.016 \\
\bottomrule
\end{tabular}

\smallskip
\begin{minipage}{0.97\textwidth}
\scriptsize
$^{a}$\texttt{CH2}: the $m_s=-1$ triplet member. $^{b}$\texttt{H3\_Linear}: the $m_s=-\tfrac12$ doublet member. These two profiles describe the specified members, not the entire subspaces.
\end{minipage}
\end{table*}

\subsection{Constructing Instances and References}
\label{sec:instance_references}

Each instance specifies a molecular geometry, basis set, and active space. Mean-field calculations provide the molecular orbitals. Active-space selection defines the electronic Hamiltonian, and the Jordan--Wigner transformation maps each active spin orbital to one qubit.

We compute exact spectra and ground states on the full $2^n$-dimensional qubit space. The calculation is not restricted to a fixed electron number or spin sector. A non-degenerate instance provides one ground-state reference. When several states share the lowest energy, we retain the full ground subspace and specify which state
supplies any reported entropy profile. Appendix~\ref{app:instances_characterization} collects the molecular settings, wire ordering, target particle numbers and reference-state conventions, together with the full descriptor definitions and values.

For cost comparisons, we also reduce circuits saved during search by removing gates and re-optimizing the remaining parameters. Only reduced circuits whose energies meet chemical accuracy serve as cost references. Section~\ref{sec:extraction} describes this procedure.

\subsection{Benchmark Instance Overview}
\label{sec:instance_overview}
\label{sec:audit}

Table~\ref{tab:criteria_audit} lists the eleven instances and their
reference properties. Near-product targets support comparisons with
compact circuits. Degenerate ground levels allow different states at
the same energy to be examined. The unique ground-state targets include
both high uniform entropy and similar mean entropies with different
local distributions.

The suite also supports two types of comparison across experimental
settings: paired connectivity tests on the same Hamiltonian, and a
\texttt{BeH2} ladder with fixed geometry and varying basis sets and
active spaces. Section~\ref{sec:regimes} defines the five diagnostic
tasks supported by these instances. Before introducing those
experiments, we define how their outputs are measured.
}
{
\section{Evaluation Framework}
\label{sec:protocol}
\label{sec:reporting}

We evaluate a returned circuit $C$ and its parameters $\theta$ along three complementary dimensions: energy accuracy and success, circuit cost, and the state actually prepared. Energy determines whether the task is solved; cost and state measurements describe differences that the success label does not capture. State diagnostics depend on the target: local entropy profiles are compared with a unique ground state, whereas degenerate targets require a ground-subspace reference.
Table~\ref{tab:reporting} summarizes the measurements and their scope.

For each method, instance, and configuration, we report the number of independent runs and the mean and best errors of their selected circuits. The best circuit has the lowest evaluation energy. Any selection across budgets or connectivity settings states that wider scope. Circuit records identify their source and parameter version; recorded rollout energies supply a separate success statistic.

\begin{table*}[t]
\centering
\footnotesize
\caption{Measurements and their scope. Circuit measurements use the selected evaluation outputs; success rates state whether they count runs or recorded rollouts.}
\label{tab:reporting}
\setlength{\tabcolsep}{4pt}
\renewcommand{\arraystretch}{1.12}
\begin{tabular}{@{}L{0.16\textwidth}L{0.22\textwidth}L{0.24\textwidth}L{0.32\textwidth}@{}}
\toprule
Measurement & Question & Applies to & Reported items \\
\midrule
Energy and success & How accurate and how often successful? & All selected outputs; the stated run or rollout records & Energy error; success-rate type, numerator, and denominator \\
\addlinespace[3pt]
Circuit cost & How costly are successful outputs? & Gate counts for all outputs; reference-relative summaries for accurate outputs with a cost reference & Gate counts; accurate-circuit count, mean cost, and relative redundancy \\
\addlinespace[3pt]
State diagnostics: unique target & Where does local entropy differ? & Outputs on non-degenerate instances, including inaccurate outputs & Full entropy profile, $\MAES$, and exact reference mean \\
\addlinespace[3pt]
State diagnostics: degenerate target & Which states were returned? & Chemically accurate outputs on degenerate instances & Ground-subspace weights, pairwise fidelities, groups, and physical labels \\
\bottomrule
\end{tabular}
\end{table*}

\subsection{Energy Accuracy and Success}
\label{sec:energy_success}

The target energy $E_0$ uses the full qubit space defined in Section~\ref{sec:instance_references}. The energy error is
\begin{equation}
\varepsilon(C,\theta)=\langle C(\theta)|H|C(\theta)\rangle-E_0 .
\label{eq:energy_error}
\end{equation}
An output reaches chemical accuracy when $\varepsilon\le1.6\,\mathrm{mHa}$. For a degenerate ground level, every member has the same target energy $E_0$.

The success rate depends on which records are counted:
\begin{itemize}
\item \emph{Run success rate:} the fraction of independent runs whose selected evaluation circuit reaches chemical accuracy.
\item \emph{Rollout success rate:} the fraction of recorded rollouts that reach chemical accuracy at the selected checkpoints. Successful and total rollout counts are pooled across runs.
\end{itemize}
Every rate states its type, numerator, and denominator. Run and rollout rates count different events and are not directly comparable as cross-method success probabilities. We report the rate alongside energy error to show both the frequency of success and the size of the remaining error.

\subsection{Circuit Cost}
\label{sec:circuit_cost}

We count primitive RX, RY, RZ, and CNOT gates before cancellation or merging. Every rotation and CNOT counts as one gate, including zero-angle rotations. Section~\ref{sec:gate_basis} describes how each method's native representation is converted to this basis.

For each instance, $C_{\mathrm{ref}}$ is the smallest chemically accurate circuit found and verified through the extraction procedure in Section~\ref{sec:extraction}.

For each method--instance--configuration with a reference, we summarize the $N$ chemically accurate selected circuits by their mean gate count and its redundancy relative to the reference:
\begin{equation}
\overline{|C|}=\frac{1}{N}\sum_{i=1}^{N}|C_i|,\qquad
\rho_{\mathrm{reported}}=1-\frac{|C_{\mathrm{ref}}|}{\overline{|C|}}.
\label{eq:mean_redundancy}
\end{equation}
For an individual accurate circuit, $\rho(C)=1-|C_{\mathrm{ref}}|/|C|$. Zero denotes the reference cost; a positive value is the fraction of gates in excess of that cost. This is a relative cost summary, not an estimate of the globally minimal circuit size. If $N=0$ or no reference is available, redundancy is not applicable. Raw gate counts can still be reported.

\subsection{State-Level Diagnostics}
\label{sec:state_diagnostics}

\subsubsection{Unique ground state: local entropy profiles}
\label{sec:local_entropy}

For a unique ground state, the entropy profile shows where an output has too much or too little single-qubit entropy. It applies to both accurate and inaccurate outputs. Using the entropy definition in
\eqref{eq:vn_entropy}, we compute
\begin{equation}
\MAES=\frac{1}{n}\sum_{k=0}^{n-1}
\left|S_{\mathrm{circuit}}(q_k)-S_{\mathrm{exact}}(q_k)\right| .
\label{eq:maes_protocol}
\end{equation}
We report this mean absolute difference with the full entropy profile and energy error. The mean summarizes the mismatch; the profile shows which qubits contribute to it. Different states can have identical local entropies, so this comparison does not establish state identity.

A pure product state has zero entropy on every qubit and therefore $\MAES=\bar S_{\mathrm{exact}}$. We report the exact mean as a reference scale for each instance. In our pure-state simulations, an output is classified as a product state within numerical tolerance when $S_{\mathrm{circuit}}(q_k)<10^{-6}$\,bit on every qubit. This test uses the full profile.

\subsubsection{Degenerate ground level: state identification}
\label{sec:subspace_identification}

For a degenerate target, the reference is the full ground subspace, since its members can have different entropy profiles. We analyze chemically accurate selected outputs, stating their configurations and sample counts. Each output's ground-subspace weight and its fidelity with another output are
\begin{equation}
p_{\mathrm{GS},i}=\langle\psi_i|P_{\mathrm{GS}}|\psi_i\rangle,\qquad
F_{ij}=\left|\langle\psi_i|\psi_j\rangle\right|^2.
\label{eq:subspace_overlap}
\end{equation}
where $P_{\mathrm{GS}}$ projects onto the exact ground subspace from Section~\ref{sec:instance_references}. We group outputs by fidelity and use observables such as $S_z$ to identify their physical differences. If no output reaches chemical accuracy, no state-identification summary is reported. The \texttt{CH2} case in Section~\ref{sec:r2_results} illustrates this analysis; it does not establish a general state-selection preference of a QAS method.

The next section connects these measurements to diagnostic experiments and describes how the methods produce the evaluation records. Appendix~\ref{app:implementation_evaluation} specifies the per-method records and ground-subspace grouping convention.
}
{
\section{Experimental Design and Procedure}
\label{sec:procedure}

We evaluate five QAS methods on the diagnostic tasks below. The comparisons use the instance references and measurement rules defined in Sections~\ref{sec:suite} and~\ref{sec:protocol}.

\subsection{Diagnostic Tasks and Instance Selection}
\label{sec:regimes}

Table~\ref{tab:criteria_audit} lists the instances and their reference properties. The five tasks define the comparisons:
\emph{R1} compares the cost of chemically accurate outputs under shallow and deep budgets against compact reference circuits.
\emph{R2} distinguishes equal-energy outputs on \texttt{CH2} by ground-subspace membership and spin.
\emph{R3} compares each output's local entropy profile with its own non-degenerate target.
\emph{R4} compares all-to-all and linear connectivity on the same Hamiltonian, holding other settings fixed within each method--instance pair.
\emph{R5} follows energy, success rates, and output states along the 8--14-qubit \texttt{BeH2} series at fixed geometry. Its basis set and active space change with register size, so it is not a controlled qubit-count-only comparison.

\subsection{Search Methods and Common Evaluation Basis}
\label{sec:gate_basis}
\label{sec:rules}

We evaluate five implementations of the approaches in Section~\ref{sec:bg_qas}: \textsc{CRLQAS}~\cite{patelcurriculum}, \textsc{HyRLQAS}~\cite{niu2025hybrid},
\textsc{QuantumDARTS}~\cite{wu2023quantumdarts}, \textsc{TFQAS}~\cite{he2024training}, and \textsc{GQEQAS}~\cite{nakaji2024generative}.

Circuits start from $\ket{0}^{\otimes n}$. Each method retains its search mechanism, with the gate-pool adaptations in Table~\ref{tab:gate_standardization}. All outputs are converted to the common evaluation basis and counted using Section~\ref{sec:circuit_cost}.

\begin{table}[t]
\centering
\scriptsize
\caption{Gate-pool standardization and evaluation-basis alignment. $\mathcal{B}=\{\mathrm{RX},\mathrm{RY},\mathrm{RZ},\mathrm{CNOT}\}$ is the common evaluation basis.}
\label{tab:gate_standardization}
\setlength{\tabcolsep}{2.5pt}
\begin{tabular}{@{}L{0.22\columnwidth}L{0.22\columnwidth}L{0.18\columnwidth}L{0.06\columnwidth}L{0.20\columnwidth}@{}}
\toprule
Method & Original pool & Search pool here & Basis & Adaptation \\
\midrule
\textsc{CRLQAS} & $\mathcal{B}$ & same & $\mathcal{B}$ & none \\
\textsc{HyRLQAS} & $\mathcal{B}$ & same & $\mathcal{B}$ & none \\
\textsc{QuantumDARTS} & \{Rz-Ry-Rz, $I$, CNOT\} & same & $\mathcal{B}$ & lowered at evaluation \\
\textsc{TFQAS} & \{RX, RY, RZ, XX, YY, ZZ\} & $\mathcal{B}$ & $\mathcal{B}$ & replaced at search \\
\textsc{GQEQAS} & operator pool & gate-token pool over $\mathcal{B}$ & $\mathcal{B}$ & replaced at search \\
\bottomrule
\end{tabular}
\end{table}

\subsection{Budgets and Experimental Controls}
\label{sec:configuration}

Each configuration fixes the method, Hamiltonian, search space, budget, and connectivity, and is repeated with independent random seeds. R1 uses shallow and deep budgets, $D=10$ and $D=50$. The budget $D$ counts gate insertions for \textsc{CRLQAS} and \textsc{HyRLQAS}, gate tokens for \textsc{GQEQAS}, moment depth for
\textsc{TFQAS}, and slot layers for \textsc{QuantumDARTS}. Equal $D$ therefore does not imply equal primitive gate counts or computational effort. Reported run counts identify the repetitions included in each comparison.

For R4, \texttt{H3\_Linear} and \texttt{H4\_Chain} are each run with all-to-all and linear-chain connectivity. Other settings are fixed within each method--instance pair. Gate counts are measured after conversion to the common basis, so the native budget $D$ and the returned gate count are reported separately.

\subsection{Output Selection and Parameter Optimization}
\label{sec:protocol_inputs}
\label{sec:reconstruction}
\label{sec:backend}

Each run supplies one selected evaluation circuit. Learning-based methods periodically sample candidates from the current model. The lowest recorded candidate energy selects the checkpoint and its output circuit for that seed. \textsc{TFQAS} evaluates a ranked candidate set once and retains its lowest-energy circuit. Selection uses energy; structural measurements are applied afterwards. Training-stage best circuits are reported separately.

The reinforcement-learning methods store the selected gate sequence but omit its optimized angles. We optimize new parameters from zero and seeded random initializations, retaining the lowest-energy set. The other methods export circuits with angles and then use the shared local optimizer.

The resulting evaluation parameters supply each circuit's energy and state measurements, including eligibility for accurate-circuit cost summaries and degenerate-state identification. Reinforcement-learning success rates instead use the recorded rollout energies at the selected checkpoints, pooled across seeds. The other
methods use run success rates from their selected evaluation circuits. Released records distinguish original rollout energies from re-optimized circuit energies; Table~\ref{tab:quantity_sources} maps reported quantities to their sources.

We use COBYLA~\cite{powell1994direct} for systems of up to six qubits and Rotosolve~\cite{ostaszewski2021structure} for larger systems. A batched GPU state-vector backend serves Rotosolve's energy evaluations. Appendix~\ref{app:implementation_evaluation} collects the implementation and training settings, candidate selection and reconstruction details, and backend timing measurements.

\subsection{Reference-Circuit Extraction}
\label{sec:extraction}

We extract compact circuits from training snapshots in a selected energy-error range. Each snapshot is re-optimized before gate importance is estimated by deleting one gate at a time and re-optimizing the remaining parameters. A beam search explores alternative deletion sequences, starting with the lowest-cost gates.

Retained circuits must satisfy an error tolerance relative to the snapshot and a bound on the error change per deletion. Only circuits that also reach chemical accuracy serve as cost references. We verify their energies and apply the appropriate state diagnostic from Section~\ref{sec:state_diagnostics}. Appendix~\ref{App:critical_structure} gives the extraction algorithm and settings, gate-importance calculation, retained gate sequences, and additional checks over one- and two-gate architectures.
}
\section{Findings}
\label{sec:baselines}
\label{sec:results}

The results show what circuit-cost and state measurements add to an energy-only comparison. We first examine cost differences among accurate outputs and physical differences among equal-energy outputs, then use local entropy profiles to describe inaccurate outputs and interpret success across the instance ladder. Each run contributes one selected evaluation-stage circuit under Section~\ref{sec:protocol_inputs}. Appendix~\ref{app:supplementary_results} collects the complete results and separate training-stage measurements.

\begin{figure*}[!t]
\centering
\includegraphics[width=\textwidth]{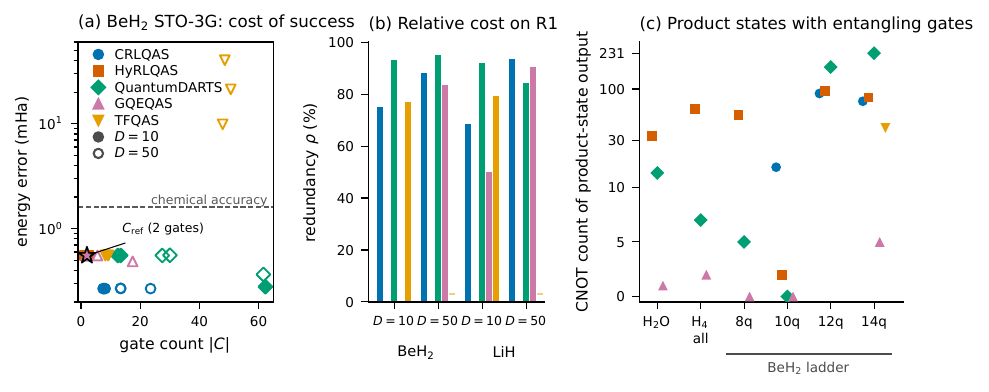}
\caption{Circuit cost and the prepared state. (a)~Energy error against gate count for every evaluation-stage circuit on \texttt{BeH2\_STO3G}, one point per seed, under the shallow (filled) and deep (open) budgets; the star marks the two-gate reference circuit. (b)~reference-relative redundancy of the mean gate count of the chemically accurate circuits on both R1 instances (Table~\ref{tab:tier1_full}); the dash marks the undefined \textsc{TFQAS} deep-budget entry. (c)~CNOT count of the best evaluation-stage circuits that Table~\ref{tab:entropy_consistency} identifies as product-state outputs ($S(q)<10^{-6}$\,bit on every qubit), on \texttt{H2O\_StrongCorr}, \texttt{H4\_Chain} (all-to-all), and the \texttt{BeH2} ladder; values from Tables~\ref{tab:eval_results} and~\ref{tab:eval_results_t5}.}
\label{fig:cost_state}
\end{figure*}

\subsection{Similar Energy Accuracy, Different Circuit Costs}
\label{sec:r1_results}

Chemically accurate outputs can differ substantially in circuit cost. Both near-product R1 targets have verified two-gate reference circuits (Fig.~\ref{fig:structure_analysis}a,b). On \texttt{BeH2\_STO3G} under the shallow budget, the best \textsc{HyRLQAS} and \textsc{GQEQAS} outputs use two rotations and no CNOT, whereas the best \textsc{QuantumDARTS} output uses 62 gates. Their energy errors differ by less than $0.3$\,mHa, despite a more than thirty-fold difference in gate count (Fig.~\ref{fig:cost_state}a).

This contrast also appears across seeds: reference-relative redundancy computed from mean gate counts ranges from $0\%$ to $93.2\%$ on shallow-budget \texttt{BeH2\_STO3G}. Increasing the budget to $D=50$ raises redundancy for three methods, while \textsc{HyRLQAS} retains two-gate circuits in every seed on both R1 instances (Fig.~\ref{fig:cost_state}b). Thus, a larger search budget does not consistently produce more economical solutions. Cost comparisons remain conditional on success; \textsc{TFQAS} has no accurate deep-budget output and hence no defined redundancy there.

Gate count also does not establish what state was prepared. Figure~\ref{fig:cost_state}c shows product-state outputs containing entangling gates, including the 14-qubit \textsc{QuantumDARTS} circuit with 231 CNOTs. These gates have entangling capability, but the evaluated circuit prepares a product state. Circuit cost therefore needs to be read alongside output-state measurements, not used as a proxy for realized entanglement.

\begin{figure}[t]
\centering
\includegraphics[width=0.95\columnwidth]{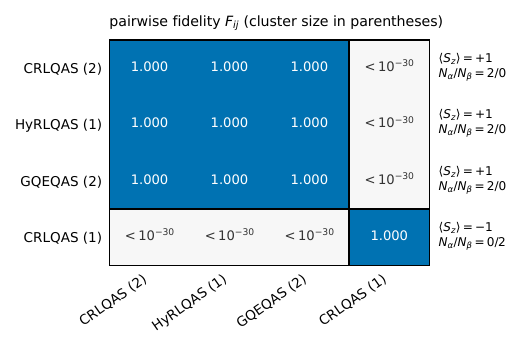}
\caption{State identification on \texttt{CH2}: pairwise fidelity between the chemically accurate evaluation-stage circuits, shown for one representative of each per-method cluster (cluster size in parentheses). The four representatives fall into two blocks, which are the $m_s=+1$ and $m_s=-1$ components of the triplet, identified by $\langle S_z\rangle$ and the spin-orbital occupations.}
\label{fig:ch2_groups}
\end{figure}

\subsection{Distinct Physical States at the Same Ground-State Energy}
\label{sec:r2_results}

On the degenerate \texttt{CH2} target, energy agreement does not imply agreement on the physical state. The six chemically accurate outputs from three methods separate into two orthogonal groups (Fig.~\ref{fig:ch2_groups}). Their ground-subspace overlaps and within-group fidelities round to $1.000$, whereas between-group fidelities are below $10^{-30}$.

Spin observables identify the groups as the $m_s=-1$ and $m_s=+1$ components of the ground-state triplet (Table~\ref{tab:ch2_spin}). Their occupations exchange spin-up and spin-down orbitals: the outputs have the same energy but different spin projections. No analyzed output occupies the $m_s=0$ component. This is an observation about the six outputs, not evidence of a general state-selection preference.

\begin{figure}[!t]
    \centering
    \begin{subfigure}{0.48\columnwidth}
        \centering
        \includegraphics[width=\linewidth]{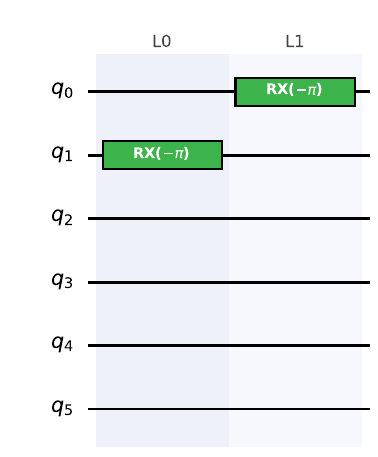}
        \caption{BeH\textsubscript{2}}
    \end{subfigure}
    \hfill
    \begin{subfigure}{0.48\columnwidth}
        \centering
        \includegraphics[width=\linewidth]{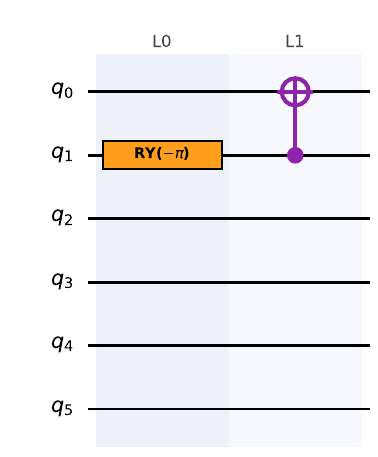}
        \caption{LiH}
    \end{subfigure}\\[4pt]
    \begin{subfigure}{0.48\columnwidth}
        \centering
        \includegraphics[width=\linewidth]{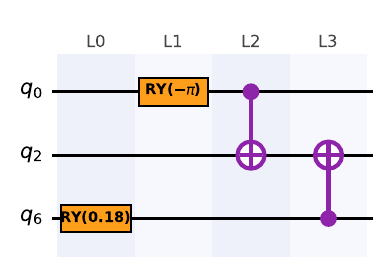}
        \caption{CH$_2$, $m_s=+1$}
    \end{subfigure}
    \hfill
    \begin{subfigure}{0.48\columnwidth}
        \centering
        \includegraphics[width=\linewidth]{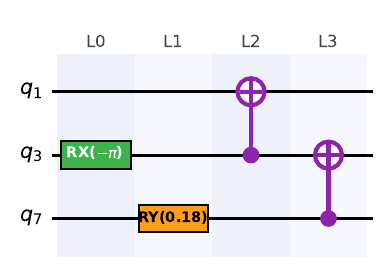}
        \caption{CH$_2$, $m_s=-1$}
    \end{subfigure}
    \caption{Reference and retained circuits. \textbf{(a,b)} Two-gate reference circuits obtained by pruning on the Regime~R1 instances: two rotations on BeH$_2$, a rotation and a CNOT on LiH; each is one member of a family of two-gate circuits with identical energy. \textbf{(c,d)} Regime~R2 yields two four-gate circuits that prepare the $m_s=+1$ and $m_s=-1$ components of the degenerate spin triplet of \texttt{CH2} (Table~\ref{tab:ch2_spin}). Qubit labels are Jordan--Wigner wires.}
    \label{fig:structure_analysis}
\end{figure}

The retained four-gate circuits make the distinction explicit: one acts on spin-up wires and the other on spin-down wires (Fig.~\ref{fig:structure_analysis}c,d). Ground-subspace and spin analysis thus resolve a physical difference that chemical accuracy alone cannot identify.

\begin{table}[t]
\centering
\footnotesize
\caption{State identification on \texttt{CH2}: the two states onto which every chemically accurate evaluation-stage circuit collapses, with the exact $m_s = 0$ component for comparison. $N_\alpha$, $N_\beta$: expected number of electrons in $\alpha$ and $\beta$ spin-orbitals. All rows have $\langle S^2\rangle = 2.000$ and energy error below $10^{-3}$\,mHa.}
\label{tab:ch2_spin}
\setlength{\tabcolsep}{5pt}
\begin{tabular}{@{}lL{0.50\columnwidth}rc@{}}
\toprule
State & Methods (cluster sizes) & $\langle S_z\rangle$ & $N_\alpha/N_\beta$ \\
\midrule
$m_s=-1$ & \textsc{CRLQAS} (1) & $-1.000$ & 0/2 \\
$m_s=+1$ & \textsc{CRLQAS} (2), \textsc{HyRLQAS} (1), \textsc{GQEQAS} (2) & $+1.000$ & 2/0 \\
$m_s=\phantom{+}0$ & none & $0.000$ & 1/1 \\
\bottomrule
\end{tabular}
\end{table}

\begin{figure*}[!t]
\centering
\includegraphics[width=\textwidth]{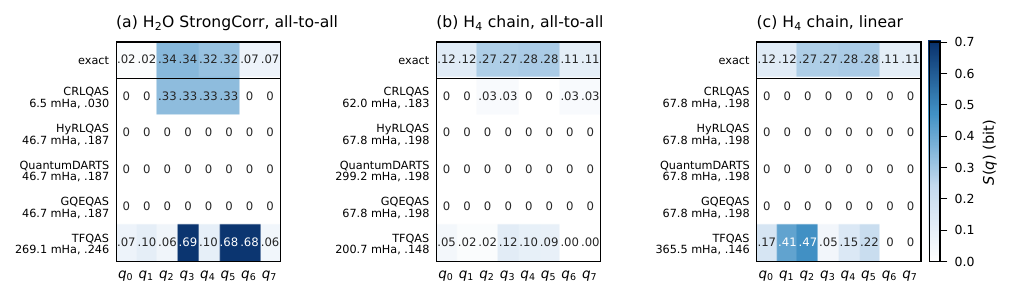}
\caption{Per-qubit entropy profiles $S(q)$ of the best evaluation-stage circuit of every method against the exact ground state, with the energy error and $\MAES$ of each circuit. (a)~\texttt{H2O\_StrongCorr}; (b)~\texttt{H4\_Chain} under all-to-all connectivity; (c)~\texttt{H4\_Chain} under linear connectivity. Rows that are zero on every qubit are product-state outputs (Section~\ref{sec:local_entropy}). Values are tabulated in Tables~\ref{tab:h2o_perqubit} and~\ref{tab:h4_perqubit_all}.}
\label{fig:profiles}
\end{figure*}

\subsection{Local Entropy Differences Between Inaccurate Outputs}
\label{sec:r3_results}
\label{sec:r4_results}

Local entropy profiles distinguish how inaccurate outputs depart from their targets. As a baseline observation, all 67 chemically accurate evaluation-stage circuits on non-degenerate instances have $\MAES\le0.013$\,bit; Table~\ref{tab:entropy_consistency} summarizes the lowest-energy outputs. The strongly entangled \texttt{H2\_Stretch} target provides a complementary check: all four methods that reach chemical accuracy reproduce its profile with $\MAES<0.001$. Neither observation makes profile agreement sufficient for energy accuracy or state identity.

On \texttt{H2O\_StrongCorr}, no method reaches chemical accuracy, but their outputs exhibit three distinct patterns (Fig.~\ref{fig:profiles}a). \textsc{CRLQAS} approximately recovers the dominant interior entropies while leaving the weaker-entropy qubits separable; its error is $6.5$\,mHa. Three methods instead return product states at $46.7$\,mHa. \textsc{TFQAS} has excess entropy on some qubits and insufficient entropy on others. The profile distinguishes these patterns without assigning a cause to the failed search.

The \texttt{H4\_Chain} results also show why entropy mismatch should not replace energy as a ranking. \textsc{TFQAS} has a smaller $\MAES$ than \textsc{CRLQAS} ($0.148$ versus $0.183$), but a larger energy error ($200.7$ versus $62.0$\,mHa). Energy determines task success; the entropy profile describes local differences from the target.

\begin{table*}[t]
\centering
\caption{Local entropy mismatch, $\MAES$ (bits), for the lowest-energy evaluation-stage circuit of every method on every non-degenerate instance, taken over the configurations of Tables~\ref{tab:eval_results} and~\ref{tab:eval_results_t5} (\texttt{H4\_Chain}: all-to-all), with $\bar{S}_{\mathrm{exact}}$ of each instance for scale. $^{*}$: that circuit reaches chemical accuracy. \texttt{CH2} is analyzed using its ground subspace (Section~\ref{sec:r2_results}); \texttt{H3\_Linear} has no chemically accurate evaluation circuit and has no eligible state-identification sample; its energy and gate counts are reported separately. \textsc{CRLQAS} and \textsc{HyRLQAS}: stored evaluation rollout with re-optimized parameters (Section~\ref{sec:reconstruction}); other methods: exported circuit. Product-state outputs, identified from the full profile by the threshold of Section~\ref{sec:local_entropy}, have $\MAES = \bar{S}_{\mathrm{exact}}$.}
\label{tab:entropy_consistency}
\small
\begin{tabular}{@{}llrrrrrrr@{}}
\toprule
Regime & Instance & $n_q$ & $\bar{S}_{\mathrm{exact}}$ & \textsc{CRLQAS} & \textsc{HyRLQAS} & \textsc{QuantumDARTS} & \textsc{GQEQAS} & \textsc{TFQAS} \\
\midrule
R1 & BeH$_2$ STO-3G & 6 & 0.006 & 0.003\rlap{$^{*}$} & 0.006\rlap{$^{*}$} & 0.003\rlap{$^{*}$} & 0.005\rlap{$^{*}$} & 0.006\rlap{$^{*}$} \\
R1 & LiH (equil.) & 6 & 0.012 & 0.003\rlap{$^{*}$} & 0.012\rlap{$^{*}$} & 0.003\rlap{$^{*}$} & 0.009\rlap{$^{*}$} & 0.009\rlap{$^{*}$} \\
R3 & H$_2$ ($R{=}2.5$\,\AA) & 4 & 0.974 & 0.000\rlap{$^{*}$} & 0.000\rlap{$^{*}$} & 0.000\rlap{$^{*}$} & 0.000\rlap{$^{*}$} & 0.026 \\
R3 & H$_2$O (stretched) & 8 & 0.187 & 0.030 & 0.187 & 0.187 & 0.187 & 0.246 \\
R3 & H$_4$ chain & 8 & 0.198 & 0.183 & 0.198 & 0.198 & 0.198 & 0.148 \\
R5 & BeH$_2$ 6-31G & 8 & 0.013 & 0.002\rlap{$^{*}$} & 0.013 & 0.013 & 0.013 & 0.188 \\
R5 & BeH$_2$ 6-311G & 10 & 0.005 & 0.005\rlap{$^{*}$} & 0.005\rlap{$^{*}$} & 0.005\rlap{$^{*}$} & 0.005\rlap{$^{*}$} & 0.401 \\
R5 & BeH$_2$ cc-pVDZ & 12 & 0.008 & 0.008 & 0.008 & 0.008 & 0.007 & 0.265 \\
R5 & BeH$_2$ cc-pVDZ & 14 & 0.019 & 0.019 & 0.019 & 0.019 & 0.019 & 0.019 \\
\bottomrule
\end{tabular}
\end{table*}

\paragraph{Connectivity as a paired comparison.}
Restricting \texttt{H4\_Chain} to linear connectivity changes the \textsc{CRLQAS} output from weakly entangled to product, while its error increases from $62.0$ to $67.8$\,mHa (Fig.~\ref{fig:profiles}b,c). For \textsc{TFQAS}, $\MAES$ barely changes ($0.148$ to $0.146$), even though the energy error rises from $200.7$ to $365.5$\,mHa. The complementary \texttt{H3\_Linear} comparison likewise shows a higher \textsc{CRLQAS} error under restricted connectivity ($9.3$ to $23.5$\,mHa), with no accurate output under either topology. These paired cases describe changes in the returned states and energies; they do not isolate whether representation, search, or parameter optimization caused the failure.

\begin{figure*}[!t]
\centering
\includegraphics[width=\textwidth]{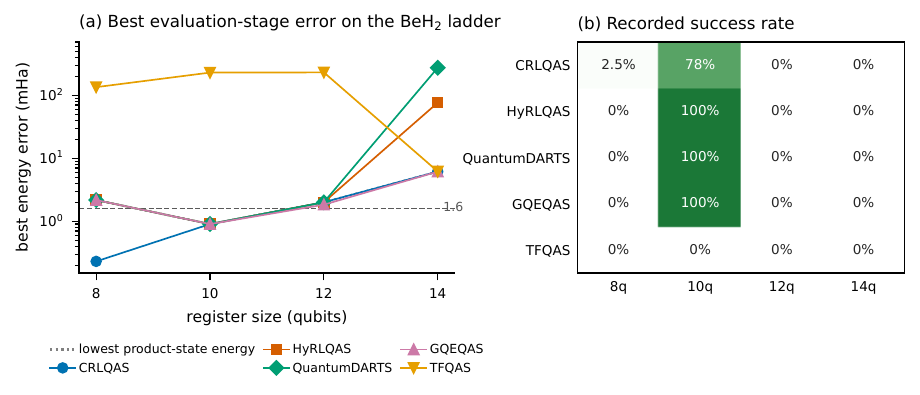}
\caption{Success on the \texttt{BeH2} ladder (Regime~R5). (a)~Best evaluation-stage energy error of every method at 8--14 qubits, with the chemical-accuracy threshold (dashed) and the lowest product-state energy found on each rung (dotted). (b)~Success rate at the selected checkpoint: rollout success rate for the reinforcement-learning methods, run success rate for the others (Table~\ref{tab:tier5_full}). These rates have different denominators and are not directly comparable across the two method groups.}
\label{fig:ladder}
\end{figure*}

\subsection{Interpreting Success Across the Instance Ladder}
\label{sec:r5_results}

The \texttt{BeH2} ladder brings energy, success rates, and output-state measurements together. Its targets remain near-product, and several methods return product states on each rung. Among the product-state outputs found, only those at 10 qubits reach chemical accuracy: the lowest errors found are $2.17$, $0.90$, $2.00$, and $6.13$\,mHa at 8, 10, 12, and 14 qubits, respectively (Fig.~\ref{fig:ladder}a).

The 10- versus 12-qubit comparison is particularly informative. At 10 qubits, \textsc{HyRLQAS} has a $100\%$ rollout success rate (40/40), while \textsc{QuantumDARTS} and \textsc{GQEQAS} have $100\%$ run success rates. At 12 qubits, all methods have zero success. Yet \textsc{CRLQAS}, \textsc{HyRLQAS}, and \textsc{QuantumDARTS} return product states at $2.00$\,mHa there. For these outputs, the change in success label reflects whether a product-state energy falls inside the target tolerance, not a change from product to entangled output.

Best error and success frequency can also move in opposite directions within one method. From 8 to 10 qubits, the best \textsc{CRLQAS} error increases from $0.23$ to $0.90$\,mHa, while its recorded rollout success rate rises from $2.5\%$ (1/40) to $78\%$ (31/40). Both statistics are needed to describe this change; neither alone summarizes performance (Fig.~\ref{fig:ladder}).

This interpretation remains instance-specific. The ladder changes basis set and active space with register size, and the product-state energies quoted are the lowest found, not certified global minima over product states. The results therefore motivate reading cross-instance success together with the target and output states, rather than attributing its variation to qubit count alone.
\section{Discussion and Limitations}
\label{sec:discussion}

\subsection{Implications for QAS Evaluation}
\label{sec:disc_evaluation}

The findings support evaluating QAS outputs through complementary questions: whether the energy target is met, at what circuit cost, and with what output-state properties. Energy remains the criterion for ground-state estimation; the additional measurements clarify differences that the success label alone leaves unresolved.

A compact, chemically accurate reference circuit provides a practical baseline for assessing excess gate cost, rather than a claim about the globally minimum circuit size. State-level diagnostics add a different perspective: entangling gates need not produce an entangled output, and equal-energy outputs need not represent the same member of a degenerate ground subspace. Reporting these properties alongside energy makes resource and state comparisons more informative without combining them into a single ranking.

Cross-instance comparisons also benefit from the target-state context. A product-state output can meet chemical accuracy on one instance and miss it on another. Success rates and energy errors should therefore be interpreted alongside the physical character of each target, rather than attributed to register size alone.

\subsection{Limitations}
\label{sec:limitations}

The benchmark is designed for controlled, instance-level comparisons. The degeneracy study covers one molecule, and the scalability ladder varies basis set and active space within one molecular family. Broader coverage would establish how widely the observed patterns recur. Exact spectral and state references support these comparisons at the present sizes; extension to larger systems would require reference approximations with quantified uncertainty.

The diagnostics have complementary scopes. Local entropy agreement does not establish full-state agreement, while degenerate targets are analyzed through their ground subspace. The reported outputs characterize the evaluated configurations and parameter-optimization protocol; they do not by themselves separate architectural restrictions from search or optimization effects.

\subsection{Future Directions}
\label{sec:disc_search}

Two extensions follow naturally. First, expanding the instance families and using controlled approximate references would broaden the benchmark's reach. Second, matched interventions could test whether the diagnostic information can improve existing QAS methods, for example by encouraging compact circuits subject to the energy target. Such studies would connect output diagnosis to search design while keeping any performance gains distinct from the evaluation evidence established here.
\section{Conclusion}
\label{sec:conclusion}

\textsc{HamQASBench} evaluates molecular ground-state QAS through energy accuracy, circuit cost, and output-state diagnostics. It combines eleven molecular Hamiltonians with exact references and a protocol for reference-relative gate cost, local entropy comparison, and state identification within degenerate ground subspaces.

Across five QAS methods, chemically accurate outputs can differ substantially in circuit cost and can represent distinct states at the same ground-state energy. Local entropy profiles distinguish inaccurate outputs and show why entangling-gate counts alone do not characterize the states prepared. The instance ladder further illustrates that chemical accuracy can be reached by product states on some targets but not by the product-state outputs found on others. Together, these findings support retaining energy as the success criterion while reporting circuit cost and state properties to make QAS comparisons more informative.

\appendices
\renewcommand{\topfraction}{0.95}
\renewcommand{\bottomfraction}{0.9}
\renewcommand{\textfraction}{0.05}
\renewcommand{\floatpagefraction}{0.7}
\renewcommand{\dbltopfraction}{0.95}
\renewcommand{\dblfloatpagefraction}{0.7}
\setcounter{topnumber}{6}
\setcounter{bottomnumber}{4}
\setcounter{totalnumber}{8}
\setcounter{dbltopnumber}{6}
\makeatletter
\setlength{\@fptop}{0pt}
\setlength{\@fpsep}{14pt plus 2pt minus 2pt}
\setlength{\@fpbot}{0pt plus 1fil}
\setlength{\@dblfptop}{0pt}
\setlength{\@dblfpsep}{14pt plus 2pt minus 2pt}
\setlength{\@dblfpbot}{0pt plus 1fil}
\makeatother
\section{Benchmark Instances and Characterization}
\label{app:instances_characterization}
\label{app:mol_details}
\label{App:Mol_details}

This appendix specifies the molecular generation settings, reference
states, and Hamiltonian descriptors for the instances in
Table~\ref{tab:criteria_audit}. Table~\ref{tab:mol_generation} lists the
generation parameters; Table~\ref{tab:hamiltonian_metadata} gives the
operator and matrix descriptors.

\subsection{Molecular Generation and Wire Ordering}
Geometries are in Angstrom (\AA). CAS$(n_e,n_o)$ specifies the active
electron and spatial-orbital counts used to generate the Hamiltonian.
The frozen-electron count is the total electron count minus the active
count. Mean-field calculations use a singlet reference
($\mathrm{mult}=1$) unless stated otherwise. Each active spin orbital
becomes one qubit under the Jordan--Wigner mapping.

\paragraph{Wire ordering.}
Wire $q_{2i}$ represents the spin-up orbital and $q_{2i+1}$ the spin-down
orbital for spatial orbital $i$. All profiles and circuit sequences use
this interleaved order. Repository circuit files use reversed simulator
labels: simulator qubit $i$ corresponds to wire $q_{n-1-i}$.

\subsection{Target Space and Reference States}
Exact diagonalization uses the full qubit space, as defined in
Section~\ref{sec:instance_references}. The column Target $N_e$ records the
particle number of the lowest eigenstate; it is distinct from the active
electron count used in Hamiltonian generation.
For \texttt{H3\_Linear}, the integrals describe the cation, but the lowest
eigenvalue lies in the three-electron sector. It is $-1.56835$\,Ha and
forms a two-fold degenerate spin doublet, $343.5$\,mHa below the lowest
two-electron state. All benchmark errors for this instance use the
doublet energy.

\paragraph{Reference members under degeneracy.}
We diagonalize $S_z$ within the exact ground subspace to select the
members whose profiles are listed in
Table~\ref{tab:criteria_audit}. For \texttt{CH2}, the reference
is the $m_s=-1$ triplet member with $\langle S^2\rangle=2$.
Both active electrons have spin down, so only spin-down wires carry
entropy. For \texttt{H3\_Linear}, the reference is the
$m_s=-\tfrac12$ doublet member with $\langle S^2\rangle=0.75$.
The opposite-spin member exchanges each pair of wires
$q_{2i}\leftrightarrow q_{2i+1}$. These member-specific profiles describe
the reference states; evaluation of degenerate outputs uses the
subspace procedure in Section~\ref{sec:subspace_identification}.

\begin{table*}[!tp]
\centering
\scriptsize
\caption{Molecular generation settings. CAS$(n_e,n_o)$ gives the active electron and spatial-orbital counts; Target $N_e$ gives the ground-state particle number. $\ddag$: H$_3^+$ integrals with a three-electron benchmark target, as explained above.}
\label{tab:mol_generation}
\setlength{\tabcolsep}{3pt}\renewcommand{\arraystretch}{1.15}
\begin{tabular}{@{}lllrlcrrr@{}}
\toprule
Regime & Instance & Geometry (\AA) & Charge & Basis & CAS$(n_e,n_o)$ & Frozen $e^-$ & Target $N_e$ & $n$ \\
\midrule
R1 & \path{BeH2_STO3G}     & Be(0,0,0); H(0,0,1.326); H(0,0,$-$1.326)        & 0 & STO-3G  & (2e, 3o) & 4 & 2 & 6  \\
R1 & \path{LiH_Equil}      & Li(0,0,0); H(0,0,1.595)                       & 0 & STO-3G  & (2e, 3o) & 2 & 2 & 6  \\
R2 & \path{CH2}             & C(0,0,0); H(0,0.86,0.73); H(0,$-$0.86,0.73)     & 0 & STO-3G  & (2e, 4o) & 6 & 2 & 8  \\
R3 & \path{H2_Stretch}     & H(0,0,0); H(0,0,2.500)                        & 0 & STO-3G  & (2e, 2o) & 0 & 2 & 4  \\
R3 & \path{H2O_StrongCorr} & O(0,0,0); H(0,1.186,0.918); H(0,$-$1.186,0.918) & 0 & STO-3G  & (4e, 4o) & 6 & 4 & 8  \\
R3, R4 & \path{H4_Chain}       & H(0,0,0); H(0,0,1); H(0,0,2); H(0,0,3)        & 0 & STO-3G  & (4e, 4o) & 0 & 4 & 8  \\
R4 & \path{H3_Linear}$^\ddag$ & H(0,0,0); H(0,0,1); H(0,0,2)               & 1 & STO-3G  & (2e, 3o) & 0 & 3 & 6  \\
R5 & \path{BeH2_631G}      & Be(0,0,0); H(0,0,1.326); H(0,0,$-$1.326)        & 0 & 6-31G   & (2e, 4o) & 4 & 2 & 8  \\
R5 & \path{BeH2_6311G}     & Be(0,0,0); H(0,0,1.326); H(0,0,$-$1.326)        & 0 & 6-311G  & (2e, 5o) & 4 & 2 & 10 \\
R5 & \path{BeH2_CCPVDZ_12q} & Be(0,0,0); H(0,0,1.326); H(0,0,$-$1.326)      & 0 & cc-pVDZ & (2e, 6o) & 4 & 2 & 12 \\
R5 & \path{BeH2_CCPVDZ_14q} & Be(0,0,0); H(0,0,1.326); H(0,0,$-$1.326)      & 0 & cc-pVDZ & (4e, 7o) & 2 & 4 & 14 \\
\bottomrule
\end{tabular}
\end{table*}
\subsection{Hamiltonian Descriptors and Computational Conventions}
\label{App:fingerprint_theory}

Spectral gaps and state profiles are defined in
Section~\ref{sec:fingerprints} and listed in
Table~\ref{tab:criteria_audit}.

\subsubsection{Descriptor Definitions}
\label{app:fp_defs}

Write the Pauli expansion as $H=c_I I+\sum_{k\in\mathcal P}c_kP_k$,
where $\mathcal P$ excludes the identity string. Let $\mathcal D$ contain
the non-identity strings made only of $I$ and $Z$, and let $w(P_k)$ be
the number of non-identity factors in $P_k$. The Pauli-weight ratios are
\begin{equation}
r_Z=\frac{\sum_{k\in\mathcal D}|c_k|}{\sum_{k\in\mathcal P}|c_k|},\qquad
r_{\ge2}=\frac{\sum_{k\in\mathcal P:\,w(P_k)\ge2}|c_k|}
                  {\sum_{k\in\mathcal P}|c_k|}.
\end{equation}
The first measures diagonal Pauli weight; the second measures weight on
strings acting on at least two qubits.

For the matrix descriptors we use the traceless Hamiltonian
$\tilde H=H-\Tr(H)I/d$, where $d=2^n$.
Write its entries as $\tilde h_{ij}$ and define
$R_i=\sum_{j\ne i}|\tilde h_{ij}|$. Then
\begin{equation}
G_1=\frac{1}{d}\sum_i\mathbf 1(|\tilde h_{ii}|\ge R_i),\qquad
G_2=\min_{R_i>0}\frac{|\tilde h_{ii}|}{R_i}.
\end{equation}
$G_1$ is the fraction of diagonally dominant rows. $G_2$ is the smallest
ratio of diagonal magnitude to off-diagonal row weight among rows with
$R_i>0$. Alternative centering rules use the same formulas on the
corresponding centered matrix.

\begin{table}[!htbp]
\centering
\footnotesize
\caption{Operator-level metadata of the eleven instances: $r_Z$, $r_{\ge2}$, $G_1$, and $G_2$ under the identity-exclusion and traceless-centering conventions specified below.}
\label{tab:hamiltonian_metadata}
\setlength{\tabcolsep}{3pt}\renewcommand{\arraystretch}{1.12}
\begin{tabular}{@{}L{0.40\columnwidth}rrrr@{}}
\toprule
Instance & $r_Z$ & $r_{\ge2}$ & $G_1$ & $G_2$ \\
\midrule
\path{BeH2_STO3G} & 0.978 & 0.593 & 0.969 & 0.094 \\
\path{LiH_Equil} & 0.876 & 0.654 & 0.891 & 0.225 \\
\path{CH2} & 0.907 & 0.483 & 0.945 & 0.210 \\
\path{H2_Stretch} & 0.728 & 0.849 & 0.750 & 0.180 \\
\path{H2O_StrongCorr} & 0.723 & 0.887 & 0.648 & 0.074 \\
\path{H4_Chain} & 0.599 & 0.809 & 0.418 & 0.011 \\
\path{H3_Linear} & 0.709 & 0.765 & 0.656 & 0.021 \\
\path{BeH2_631G} & 0.953 & 0.560 & 0.980 & 0.046 \\
\path{BeH2_6311G} & 0.923 & 0.533 & 0.963 & 0.010 \\
\path{BeH2_CCPVDZ_12q} & 0.889 & 0.498 & 0.904 & 0.019 \\
\path{BeH2_CCPVDZ_14q} & 0.767 & 0.700 & 0.621 & 0.000 \\
\bottomrule
\end{tabular}
\end{table}

\subsubsection{Computational Conventions}
\label{app:conventions}
\label{app:conv_used}

\paragraph{Identity exclusion.}
The identity coefficient includes nuclear-repulsion and frozen-core
contributions. On \texttt{BeH2\_STO3G}, it accounts for $85.5\%$ of the
absolute coefficient sum when the identity is included. Including it
would raise $r_Z$ from $0.978$ to $0.997$.
Both Pauli ratios above exclude this term.

\paragraph{Traceless centering.}
Adding a constant multiple of the identity shifts all energies equally
and leaves $\tilde H$ unchanged. All four reported descriptors are thus
invariant to this shift. Applying $G_1$ and $G_2$ to unshifted $H$ instead
can change their values substantially. On \texttt{BeH2\_STO3G}, it changes
$G_1$ from $0.969$ to $1.000$ and $G_2$ from $0.094$ to $377$.

\paragraph{Alternative conventions.}
\label{app:conv_sensitivity}
Table~\ref{tab:convention_sensitivity} compares three centering rules.
A change of center moves the diagonal entries while leaving $R_i$ fixed.
For example, setting the smallest diagonal entry to zero raises $G_1$
on \texttt{H4\_Chain} from $0.418$ to $0.887$.
The table uses the full qubit space. Restricting the particle-number
sector is a separate change of matrix representation: under the
unshifted rule, the $256$- to $70$-dimensional restriction on
\texttt{H4\_Chain} changes $G_1$ from $0.488$ to $0.500$.
On \texttt{BeH2\_STO3G}, the $64$- to $15$-dimensional restriction changes
$G_2$ from $377.0$ to $380.1$.

\begin{table}[!htbp]
\centering
\caption{Diagonal-dominance metadata under three centering rules. The
traceless rule is used throughout this article; all three rows of a block
describe the same physical Hamiltonian.}
\label{tab:convention_sensitivity}
\footnotesize
\setlength{\tabcolsep}{3pt}\renewcommand{\arraystretch}{1.12}
\begin{tabular}{@{}L{0.31\columnwidth}L{0.37\columnwidth}rr@{}}
\toprule
Instance & Centering rule & $G_1$ & $G_2$ \\
\midrule
\path{BeH2_STO3G} & traceless (used here) & 0.969 & 0.094 \\
                     & smallest diagonal $=0$ & 0.984 & 0.000 \\
                     & unshifted $H$          & 1.000 & 377.0 \\
\addlinespace
\path{H2_Stretch} & traceless (used here) & 0.750 & 0.180 \\
                     & smallest diagonal $=0$ & 0.938 & 0.810 \\
                     & unshifted $H$          & 1.000 & 2.106 \\
\addlinespace
\path{H4_Chain}   & traceless (used here) & 0.418 & 0.011 \\
                     & smallest diagonal $=0$ & 0.887 & 0.000 \\
                     & unshifted $H$          & 0.488 & 0.035 \\
\bottomrule
\end{tabular}
\end{table}

\subsubsection{Relation Between Pauli Weights and Matrix Entries}

Only strings in $\mathcal D$ contribute to the diagonal of $\tilde H$.
Their signed contributions can cancel within a row.
Each string outside $\mathcal D$ contains $X$ or $Y$ and maps a
computational-basis vector to a different basis vector, up to a phase.
The triangle inequality therefore gives
\begin{equation}
R_i\le\sum_{k\in\mathcal P\setminus\mathcal D}|c_k|.
\end{equation}
The diagonal coefficient weight supplies no corresponding lower bound
on $|\tilde h_{ii}|$. We compute the row-wise descriptors from the
matrix rather than infer them from $r_Z$.
\section{Implementation and Evaluation Details}
\label{Appendix:Method}
\label{app:implementation_evaluation}

This appendix collects the method implementations, training settings,
evaluation records, and execution backend used in
Section~\ref{sec:procedure}.

\subsection{Search Mechanisms}
\label{App:mechanisms}

Gate-pool adaptations are summarized in
Table~\ref{tab:gate_standardization}; native budget units are defined in
Section~\ref{sec:configuration}.

\paragraph{CRLQAS~\cite{patelcurriculum}.} \textsc{CRLQAS} formulates QAS as a Markov decision process in which a DQN~\cite{van2016deep} agent sequentially inserts gates into a growing circuit. The step reward is defined as
\begin{equation}
r_t =
\begin{cases}
+5 & \text{if } E_t < \xi_t \\
-5 & \text{if } t = L-1 \text{ and } E_t \geq \xi_t \\
\mathrm{clip}\!\left(\frac{E_{t-1} - E_t}{|E_{t-1} - E_{\min}| + \epsilon},\,-1,\,1\right) & \text{otherwise,}
\end{cases}
\end{equation}
where $\xi_t$ is the acceptance threshold. Depending on the task, the threshold is fixed or progressively tightened by a curriculum. Training uses $n$-step DQN ($n=6$) with experience replay. An episode ends when the circuit reaches the threshold or maximum depth $D$.

\paragraph{HyRLQAS~\cite{niu2025hybrid}.}
\textsc{HyRLQAS} predicts a gate placement and initial rotation angle at each step. These angles initialize the external optimizer. A refinement mechanism also updates previously assigned angles after each insertion. Training uses REINFORCE~\cite{williams1992simple}, with the same curriculum and stopping rules as \textsc{CRLQAS}.

\paragraph{QuantumDARTS~\cite{wu2023quantumdarts}.}
\textsc{QuantumDARTS} assigns continuous architecture weights to $\{\mathrm{Rz\text{-}Ry\text{-}Rz}, I, \mathrm{CNOT}\}$ and jointly optimizes these weights and gate parameters through a Gumbel-softmax relaxation. After search, the architecture is discretized and its parameters are re-optimized with the shared local optimizer. Identity choices leave slots empty. At evaluation, each $\mathrm{Rz\text{-}Ry\text{-}Rz}$ block is expanded into three primitive rotations, $\{\mathrm{RZ}, \mathrm{RY}, \mathrm{RZ}\}$.

\paragraph{TFQAS~\cite{he2024training}.}
\textsc{TFQAS} filters candidates in three stages. It first ranks $S$ sampled circuits by DAG path count~\cite{nam2018automated}, then re-ranks the top $R$ by expressibility~\cite{sim2019expressibility}, and finally optimizes the top $K$ with the shared local optimizer. Circuits are sampled gate by gate. Any gate that would exceed moment depth $D$ is rejected and resampled. For this benchmark, CNOT replaces the original entangler set $\{\mathrm{XX}, \mathrm{YY}, \mathrm{ZZ}\}$ during search.

\paragraph{GQEQAS~\cite{nakaji2024generative}.} 
\textsc{GQEQAS} trains an autoregressive transformer online. At each epoch, it generates circuits, evaluates their energies through VQE, and updates the model with an energy-weighted gradient. Sequence lengths are sampled uniformly up to the budget. The exported circuit is re-optimized with the shared local optimizer. For this benchmark, a discretized token pool over $\{\mathrm{RX}, \mathrm{RY}, \mathrm{RZ}, \mathrm{CNOT}\}$ replaces the original operator pool.

\subsection{Training Configuration and Budgets}
\label{Appendix:Setup}

\paragraph{Per-method training configuration.}
Table~\ref{tab:hyperparams} summarizes the training settings. The training units and budget conventions are detailed below.

\textit{CRLQAS and HyRLQAS.} One episode is a circuit-construction trajectory of at most $D$ gate insertions, with 8 environments rolled out in parallel. The two methods share the same configuration except for the batch size, which follows their update rules (DQN replay for \textsc{CRLQAS}, REINFORCE for \textsc{HyRLQAS}).

\textit{QuantumDARTS.} The search phase runs for $n_\mathrm{epochs}$ with $n_\mathrm{inner}=10$ gradient steps per epoch. Architecture weights are updated only during this phase; afterwards the best discrete circuit undergoes fixed-structure parameter re-optimization with COBYLA or Rotosolve.

\textit{TFQAS.} No gradient updates are performed; the budget is the number of candidate circuits sampled.

\textit{GQEQAS.} One epoch draws \texttt{online\_sample\_count}$=250$ circuits from the current generative model, evaluates them via VQE, and updates the transformer with the energy-weighted objective.

\begin{table*}[!tp]
\centering
\footnotesize
\caption{Key training hyperparameters per method. $D$ is the budget of Section~\ref{sec:configuration} in each method's native unit (shallow: $D=10$; deep: $D=50$); its meaning per method is given in the text. VQE optimizer: COBYLA (\texttt{global\_iters}$=100$) for $\le6$ qubits; Rotosolve (\texttt{sweeps}$=2$) for $>6$ qubits. Episode counts are given as easy/hard by molecule complexity.}
\label{tab:hyperparams}
\setlength{\tabcolsep}{6pt}
\begin{tabular}{@{}llllll@{}}
\toprule
Method & Training unit & Budget & $D$ & Policy update & Batch \\
\midrule
\textsc{CRLQAS} & episode ($\times$8 envs) & 8k / 15k ep & 10 / 50 & DQN ($n$-step, $n=6$) & 1000 \\
\textsc{HyRLQAS} & episode ($\times$8 envs) & 8k / 15k ep & 10 / 50 & REINFORCE & 16 \\
\textsc{QuantumDARTS} & epoch ($n_\mathrm{inner}=10$) & 500--900 ep & 10 / 50 & Adam ($\theta$: $10^{-2}$--$5\times 10^{-3}$; $\alpha$: $5\times 10^{-3}$--$3\times 10^{-3}$) & --- \\
\textsc{TFQAS} & candidate circuits & $S=11$k--23k & 10 / 50 & none & --- \\
\textsc{GQEQAS} & epoch (online refresh) & 1k / 1.5k ep & 10 / 50 & Adam & --- \\
\bottomrule
\end{tabular}
\end{table*}

\paragraph{Budget enforcement.}
Circuits may use less than the full budget. The reinforcement-learning
methods stop at the acceptance threshold, \textsc{QuantumDARTS} can
select identity operations, \textsc{GQEQAS} can select an end-of-sequence
token, and \textsc{TFQAS} samples circuits of different sizes.

\textsc{QuantumDARTS} search includes an expected-gate-cost penalty with weight $0.003$. The released configurations also record a compiled-count flag for diagnostics; this flag does not affect selection.

Table~\ref{tab:budgets} gives the per-instance budgets. Outside Regime~R1, the four methods with gate- or depth-count budgets use the same $D$ on each instance, except on \texttt{H3\_Linear}: \textsc{CRLQAS} uses $D=70$, and the others use $D=50$. \textsc{QuantumDARTS} uses $D=10$ and $50$ slot layers in Regime~R1 and $D=n+4$ slot layers on the other $n$-qubit instances.

\begin{table}[!htbp]
\centering
\footnotesize
\caption{Budget $D$ per instance for \textsc{CRLQAS}, \textsc{HyRLQAS} (gate insertions), \textsc{GQEQAS} (tokens), and \textsc{TFQAS} (moment depth). $^{a}$\textsc{CRLQAS} uses $D=70$; the other three use $D=50$. \texttt{CH2} was additionally run under $D=20$ for the two reinforcement-learning methods.}
\label{tab:budgets}
\setlength{\tabcolsep}{4pt}
\begin{tabular}{@{}L{0.36\columnwidth}r@{\hspace{10pt}}L{0.36\columnwidth}r@{}}
\toprule
Instance & $D$ & Instance & $D$ \\
\midrule
\texttt{BeH2\_STO3G}, \texttt{LiH\_Equil} & 10 / 50 & \texttt{H3\_Linear} (both connectivities) & 50$^{a}$ \\
\texttt{CH2} & 70 (20) & \texttt{BeH2\_631G} & 70 \\
\texttt{H2\_Stretch} & 50 & \texttt{BeH2\_6311G} & 100 \\
\texttt{H2O\_StrongCorr} & 70 & \texttt{BeH2\_CCPVDZ\_12q} & 100 \\
\texttt{H4\_Chain} (both connectivities) & 70 & \texttt{BeH2\_CCPVDZ\_14q} & 100 \\
\bottomrule
\end{tabular}
\end{table}

\paragraph{Training budgets.}
Training units have different costs across methods. An RL episode can require up to $D$ sequential VQE calls, whereas a \textsc{QuantumDARTS} epoch processes all circuit slots in a single backward pass. We report budgets in each method's native units. Table~\ref{tab:optimizer_speedup} measures backend speed-up on a representative \textsc{CRLQAS} workload.
\subsection{Evaluation Records and Circuit Reconstruction}
\label{App:eval_records}

\subsubsection{Candidate Evaluation and Checkpoint Selection}

Table~\ref{tab:eval_records} gives the candidate counts and stored
records. Learning-based methods evaluate checkpoints without gradient
updates. For each seed, the lowest recorded candidate energy selects
the checkpoint and its circuit. \textsc{TFQAS} selects the lowest-energy
circuit from its single optimized candidate set.

\begin{table}[!htbp]
\centering
\scriptsize
\caption{Evaluation records per method: what is evaluated at each checkpoint, how the evaluation-stage circuit is selected, and what the stored record contains.}
\label{tab:eval_records}
\setlength{\tabcolsep}{3pt}
\begin{tabular}{@{}L{0.23\columnwidth}L{0.28\columnwidth}L{0.19\columnwidth}L{0.22\columnwidth}@{}}
\toprule
Method & Evaluated per checkpoint & Selection & Stored record \\
\midrule
\textsc{CRLQAS}, \textsc{HyRLQAS} & $K_{\mathrm{eval}}=20$ policy rollouts every $e_{\mathrm{eval}}$ episodes; energies recorded & checkpoint with the lowest recorded energy, per seed & gate sequence of the selected rollout; rollout energies \\
\addlinespace[2pt]
\textsc{QuantumDARTS} & 24--30 discretized circuits (instance-dependent) & as above & circuit with angles \\
\addlinespace[2pt]
\textsc{GQEQAS} & 50 generated sequences & as above & circuit with angles \\
\addlinespace[2pt]
\textsc{TFQAS} & one pass: top $K=50$ ranked candidates optimized & lowest-energy candidate & circuit with angles \\
\bottomrule
\end{tabular}
\end{table}

\subsubsection{Stored Records and Parameter Optimization}

The reinforcement-learning methods retain the gate sequence of the
selected rollout and the checkpoint's rollout energies, but omit the
optimized angles. We optimize new parameters for that gate sequence
from zero and from four random restarts, reduced to two at 12 qubits or
more, and retain the lowest energy
(Section~\ref{sec:reconstruction}). This procedure supplies an evaluated
state for the selected architecture; it does not recover the original
rollout parameters. The other methods export circuits with angles and
use the shared optimizer as stated in Table~\ref{tab:quantity_sources}.

\subsubsection{Measurement Sources and Success Rates}

Table~\ref{tab:quantity_sources} identifies which record supplies each
measurement. Reinforcement-learning records retain both the original
rollout energy and the re-optimized circuit energy. Each seed contributes
20 recorded rollouts to the pooled success rate; the other methods
contribute one selected circuit per seed to the run success rate.
These rates count different events and cannot support a direct
cross-method ranking of success probabilities. Aggregation follows
Section~\ref{sec:reporting}.

\begin{table}[!htbp]
\centering
\footnotesize
\caption{Record that supplies each reported quantity (Section~\ref{sec:reconstruction}).}
\label{tab:quantity_sources}
\setlength{\tabcolsep}{4pt}
\begin{tabular}{@{}L{0.30\columnwidth}L{0.64\columnwidth}@{}}
\toprule
Quantity & Source \\
\midrule
Best energy error & re-optimized evaluation circuit (reinforcement-learning methods); exported circuit after re-optimization with the shared optimizer (others) \\
Success rate & rollout success rate: recorded rollout energies at the selected checkpoint, pooled over seeds (reinforcement-learning methods); run success rate: fraction of seeds whose evaluation-stage circuit is chemically accurate (others) \\
Gate count; $\rho$ & selected evaluation circuit; cost summaries include only circuits whose evaluation energy reaches chemical accuracy \\
Entropy profile; $\MAES$; state identification & state prepared with the evaluation parameters; entropy comparison includes inaccurate outputs on non-degenerate instances; degenerate-state identification analyzes accurate outputs \\
Training-stage best & lowest-error training snapshot of the two reinforcement-learning methods, reported separately (Section~\ref{sec:protocol_inputs}; Appendix~\ref{App:training_results}) \\
\bottomrule
\end{tabular}
\end{table}

\subsubsection{Ground-Subspace Comparison and Grouping}
\label{app:state_grouping}

For an orthonormal basis $\{|g_a\rangle\}_{a=1}^{d_{\mathrm{GS}}}$ of
the exact ground subspace, its projector is
$P_{\mathrm{GS}}=\sum_a|g_a\rangle\langle g_a|$.
The overlap in \eqref{eq:subspace_overlap} therefore sums the squared
overlaps with all basis members and is independent of the chosen basis.

Single-linkage clustering at $F_{ij}\ge0.9$ forms connected components
of a graph whose vertices are the analyzed outputs. An edge joins two
outputs when their fidelity reaches the threshold. The threshold applies
to edges, not necessarily to every pair within a connected component.
Cluster populations count all included circuits, even when a figure
shows only representatives. Between-cluster fidelities refer to pairs
of outputs in different components. Section~\ref{sec:r2_results} reports
the pairwise values and populations for the \texttt{CH2} analysis, then
uses spin observables to label its groups.
\subsection{Execution Backend}
\label{App:backend}

\paragraph{Execution.} The backend groups energy evaluations to make use of GPU parallelism. Rotosolve evaluates three angles, $\{0,\pi/2,\pi\}$, for each parameter and computes its update analytically. We batch these energy queries and overlap $k=10$ optimization environments across CUDA streams. COBYLA instead issues adaptive sequential queries, with little opportunity for grouped evaluation. We use it only on systems of at most six qubits.

\paragraph{Timing.} Table~\ref{tab:optimizer_speedup} compares four optimizers on CPU, GPU, and grouped GPU execution. Each entry measures one training episode of a representative \textsc{CRLQAS} workload on the \texttt{BeH2} ladder. Rotosolve uses two complete sweeps per optimization call (\texttt{rotosolve\_sweeps}$=2$); COBYLA uses \texttt{global\_iters}$=100$.

\begin{table}[!htbp]
\centering
\footnotesize
\caption{Wall-clock time per training episode (seconds) on a representative \textsc{CRLQAS} workload over the \texttt{BeH2} basis-set ladder. GPU$^*$ denotes grouped GPU execution with $k=10$ parallel environments.}
\label{tab:optimizer_speedup}
\setlength{\tabcolsep}{5pt}
\begin{tabular}{@{}llrrrr@{}}
\toprule
Hardware & Optimizer & 8q & 10q & 12q & 14q \\
\midrule
CPU & COBYLA & 48.9 & 146.7 & 270.9 & 429.9 \\
 & Rotosolve & 42.7 & 144.9 & 195.7 & 338.0 \\
 & AdamSPSA & 69.7 & 218.4 & 439.1 & 934.2 \\
 & PSRAdam & 54.5 & 206.4 & 243.4 & 443.0 \\
\midrule
GPU & COBYLA & --- & --- & --- & --- \\
 & Rotosolve & 19.0 & 16.2 & 21.1 & 63.2 \\
 & AdamSPSA & 95.6 & 98.3 & 110.9 & 218.8 \\
 & PSRAdam & 45.2 & 55.8 & 69.2 & 131.1 \\
\midrule
GPU$^*$ & COBYLA & --- & --- & --- & --- \\
 & Rotosolve & 14.7 & 12.9 & 13.4 & 27.0 \\
 & AdamSPSA & 79.0 & 70.7 & 81.3 & 104.2 \\
 & PSRAdam & 40.9 & 48.7 & 48.4 & 69.8 \\
\bottomrule
\end{tabular}
\end{table}
\section{Reference-Circuit Extraction}
\label{App:critical_structure}

Algorithm~\ref{alg:critical_structure} extracts compact circuits from low-error training snapshots for a fixed molecule and method. It returns one retained circuit per snapshot and summarizes recurring gate patterns across snapshots. Table~\ref{tab:prune_settings} lists the settings.

\subsection{Extraction Algorithm and Settings}

\begin{table}[!htbp]
\centering
\footnotesize
\caption{Settings of the extraction procedure.}
\label{tab:prune_settings}
\setlength{\tabcolsep}{5pt}
\begin{tabular}{@{}L{0.14\columnwidth}L{0.43\columnwidth}L{0.34\columnwidth}@{}}
\toprule
Symbol & Meaning & Value \\
\midrule
$w$ & error-bucket width & 0.01\,mHa \\
$\delta_{\mathrm{bucket}}$ & bucket slack & 0.05\,mHa \\
$\delta_{\mathrm{rec}}$ & reconstruction slack & 0.3\,mHa \\
$\tau$ & per-step change bound & 0.2\,mHa \\
$B$ & beam width & 4 \\
$k$ & branching factor & 3 \\
$M$ & counterfactual evaluations per snapshot & 1000 \\
--- & anchor signatures & 3 \\
--- & anchor cost factor & $1/0.15$ \\
--- & inner re-optimization & COBYLA, 300 it., 2 restarts \\
\bottomrule
\end{tabular}
\end{table}

\begin{algorithm}[t]
\caption{Reference-Circuit Extraction}
\label{alg:critical_structure}
\begin{algorithmic}[1]
\Require Snapshot records $\mathcal{S}$, bucket width $w$, selected bucket $b$, beam width $B$, branching factor $k$, evaluation budget $M$, bucket slack $\delta_{\mathrm{bucket}}$, reconstruction slack $\delta_{\mathrm{rec}}$, per-step change bound $\tau$
\Ensure Retained circuits $\{C_{\mathrm{ret}}\}$ and cross-snapshot structure 
summaries

\State Bucket snapshot events by quantized error; select records $\mathcal{S}_b$ in bucket $b$
\State Identify anchor signatures $\mathcal{A}$ from frequent gate-qubit signatures in $\mathcal{S}_b$
\State Sample representative subset $\widetilde{\mathcal{S}}_b \subseteq 
\mathcal{S}_b$, balanced across runs

\For{each snapshot $s \in \widetilde{\mathcal{S}}_b$}
    \State Reconstruct circuit $C$ and re-optimize parameters to obtain baseline 
    $C^{(0)}$ with error $\varepsilon(C^{(0)})$
    \State Set $\varepsilon_{\mathrm{allow}} = \max(b + \delta_{\mathrm{bucket}},\ 
    \varepsilon(C^{(0)}) + \delta_{\mathrm{rec}})$

    \For{each gate $g_j \in C^{(0)}$}
        \State Remove $g_j$, re-optimize from warm start, compute 
        $\Delta_e(j) = \varepsilon(C^{(0)} \setminus g_j) - \varepsilon(C^{(0)})$
    \EndFor
    \State Set deletion cost $c(j) = |\Delta_e(j)|$; for gates in
    $\mathcal{A}$, divide $c(j)$ by the anchor factor $0.15$

    \State Initialize $\mathcal{B}_0 = \{C^{(0)}\}$, evaluation counter 
    $m \leftarrow |C^{(0)}|$

    \For{$t = 0, 1, \dots$ until no feasible expansion or $m \ge M$}
        \State $\mathcal{B}_{t+1} \leftarrow \emptyset$
        \For{each $C \in \mathcal{B}_t$}
            \State Select the $k$ gates of $C$ with lowest cost $c(j)$
            \For{each selected gate $g_j$}
                \State $C' \leftarrow C \setminus g_j$; re-optimize from warm 
                start; $m \leftarrow m + 1$
                \If{$\varepsilon(C') \le \varepsilon_{\mathrm{allow}}$ \textbf{and} 
                $|\varepsilon(C') - \varepsilon(C)| \le \tau$}
                    \State Add $C'$ to $\mathcal{B}_{t+1}$
                \EndIf
            \EndFor
        \EndFor
        \State Deduplicate $\mathcal{B}_{t+1}$ by gate sequence; retain top $B$ 
        ranked by $(|C|,\ \varepsilon(C),\ \Delta_{\mathrm{total}})$
    \EndFor
    \State $C_{\mathrm{ret}} \leftarrow$ smallest feasible circuit encountered
\EndFor
\State Aggregate retained structures across snapshots
\end{algorithmic}
\end{algorithm}

Algorithm~\ref{alg:critical_structure} gives the deletion and acceptance
rules. Snapshot sampling favours late training and balances random seeds.
Gate-importance evaluations start from the baseline parameters, and
subsequent deletions use warm starts. Deletion costs are computed once
on the baseline and remain fixed during pruning. The tie-breaker
$\Delta_{\mathrm{total}}$ is the cumulative error change.

\subsection{Retained Outputs and Cost References}

\paragraph{Outputs and counts.}
For each snapshot, we record the retained circuit $C_{\mathrm{ret}}$ and pruning ratio $r_{\mathrm{prune}}=(|C^{(0)}|-|C_{\mathrm{ret}}|)/|C^{(0)}|$. This ratio measures the reduction from the snapshot baseline; reference-relative redundancy $\rho$ instead compares a search output with a reference circuit. We also aggregate retained gate--qubit signatures across snapshots.

The algorithm's feasibility bounds are relative to each snapshot's
error. After pruning, the chemical-accuracy test in
Section~\ref{sec:circuit_cost} determines which retained circuits can serve
as $C_{\mathrm{ref}}$.

\paragraph{Scope of the cost reference.}
The reference is the smallest chemically accurate circuit found in the
stated evaluation basis. If $m$ is the true minimum under the same
circuit conventions, then $m\le|C_{\mathrm{ref}}|$ and
$1-|C_{\mathrm{ref}}|/|C|\le1-m/|C|$.
The reported redundancy is therefore a lower bound on redundancy
relative to that minimum. The finite optimization checks below do not
establish global optimality.

\subsection{Circuit Verification Records}

\begin{table}[!htbp]
\centering
\footnotesize
\caption{Outcome of the extraction on the Regime~R1 instances: retained size by snapshot error level, and the exhaustive check over all one- and two-gate architectures in the common basis.}
\label{tab:extraction_outcomes}
\setlength{\tabcolsep}{4pt}
\begin{tabular}{@{}L{0.34\columnwidth}L{0.30\columnwidth}L{0.30\columnwidth}@{}}
\toprule
 & \texttt{BeH2\_STO3G} & \texttt{LiH\_Equil} \\
\midrule
Snapshots at the product-state error & 0.554\,mHa $\to$ 2 gates & 1.054\,mHa $\to$ 2 gates \\
\addlinespace[2pt]
Lower-error snapshots & 0.268\,mHa $\to$ 6--12 gates & 0.228\,mHa $\to$ 6--12 gates \\
\addlinespace[2pt]
Best of 48 one-gate architectures & 423\,mHa & 287\,mHa \\
\addlinespace[2pt]
Best of 2304 two-gate architectures & 0.554\,mHa (reference) & 1.054\,mHa (reference) \\
\bottomrule
\end{tabular}
\end{table}

\paragraph{Regime R1 reference circuits.}
Table~\ref{tab:cref_sequences} lists the two Regime~R1 reference circuits, drawn in Fig.~\ref{fig:structure_analysis}a,b; Table~\ref{tab:extraction_outcomes} summarizes the extraction outcomes. Both occupy wires $q_0$ and $q_1$, the spin orbitals of the lowest active orbital (Appendix~\ref{app:mol_details}). Each is one of several equal-energy two-gate circuits: either two $\pm\pi$ rotations about $x$ or $y$, or one such rotation followed by a CNOT. Different snapshots yield different members of this family.

Table~\ref{tab:extraction_outcomes} compares the retained sizes at two
snapshot error levels. We also optimized all one-gate and ordered
two-gate architectures in the common basis using the same
re-optimization procedure. No one-gate candidate reached chemical
accuracy, and no two-gate candidate improved on the references.

\begin{table}[!htbp]
\centering
\footnotesize
\caption{Reference and retained circuits obtained by pruning, acting on $\ket{0}^{\otimes n}$; $q_k$ denotes wire $k$. Errors are re-optimized energy errors.}
\label{tab:cref_sequences}
\setlength{\tabcolsep}{4pt}
\begin{tabular}{@{}L{0.22\columnwidth}L{0.56\columnwidth}r@{}}
\toprule
Instance & Gate sequence & Error (mHa) \\
\midrule
\texttt{BeH2\_STO3G} & $\mathrm{RX}(-\pi)\,q_1$; $\mathrm{RX}(-\pi)\,q_0$ & 0.554 \\
\texttt{LiH\_Equil} & $\mathrm{RY}(-\pi)\,q_1$; $\mathrm{CNOT}(q_1\!\to\!q_0)$ & 1.054 \\
\addlinespace[2pt]
\texttt{CH2}, $m_s=+1$ & $\mathrm{RY}(0.180)\,q_6$; $\mathrm{RY}(-\pi)\,q_0$; \newline $\mathrm{CNOT}(q_0\!\to\!q_2)$; $\mathrm{CNOT}(q_6\!\to\!q_2)$ & $<$0.001 \\
\addlinespace[2pt]
\texttt{CH2}, $m_s=-1$ & $\mathrm{RX}(-\pi)\,q_3$; $\mathrm{RY}(0.180)\,q_7$; \newline $\mathrm{CNOT}(q_3\!\to\!q_1)$; $\mathrm{CNOT}(q_7\!\to\!q_3)$ & $<$0.001 \\
\bottomrule
\end{tabular}
\end{table}

\paragraph{Regime R2 retained circuits.}
On \texttt{CH2}, the smallest retained circuits have four gates (Table~\ref{tab:cref_sequences}). The $m_s=+1$ circuit occupies spin-up wires only and the $m_s=-1$ circuit spin-down wires only; their entropy profiles ($0.068$\,bit on two wires, zero elsewhere) coincide with those of the exact members (Table~\ref{tab:criteria_audit}). Fig.~\ref{fig:structure_analysis}c,d draws both circuits.
\section{Supplementary Results}
\label{app:supplementary_results}
\label{App:full_results}

\subsection{Best Evaluation-Stage Circuits}
\label{App:eval_tables}

Tables~\ref{tab:eval_results} and~\ref{tab:eval_results_t5} report the best evaluation-stage circuit of every method on every instance, with its energy error, CNOT count, and rotation count. Here, best means the lowest evaluation energy among selected per-run circuits; column headings show the budget and connectivity splits. Table~\ref{tab:tier1_redundancy} gives the reference-relative redundancy on \texttt{BeH2\_STO3G} drawn in Fig.~\ref{fig:cost_state}b, and Tables~\ref{tab:h2o_perqubit} and~\ref{tab:h4_perqubit_all} the per-qubit profiles drawn in Fig.~\ref{fig:profiles}.

\begin{table*}[!tp]
\centering
\scriptsize
\caption{Best evaluation-stage circuit of every method on regimes R1--R4: energy error (mHa), CNOT count, and rotation-gate count. Provenance as in Table~\ref{tab:entropy_consistency}; errors below $0.001$\,mHa are shown as $<$0.001.}
\label{tab:eval_results}
\setlength{\tabcolsep}{4pt}
\begin{tabular}{@{}lrrrrrrrrrrrrrrr@{}}
\toprule
Method & \multicolumn{3}{c}{\texttt{BeH2\_STO3G}} & \multicolumn{3}{c}{\texttt{BeH2\_STO3G}} & \multicolumn{3}{c}{\texttt{LiH\_Equil}} & \multicolumn{3}{c}{\texttt{LiH\_Equil}} & \multicolumn{3}{c}{\texttt{CH2}} \\
 & \multicolumn{3}{c}{$D=10$} & \multicolumn{3}{c}{$D=50$} & \multicolumn{3}{c}{$D=10$} & \multicolumn{3}{c}{$D=50$} & \multicolumn{3}{c}{} \\
\cmidrule(lr){2-4} \cmidrule(lr){5-7} \cmidrule(lr){8-10} \cmidrule(lr){11-13} \cmidrule(lr){14-16}
 & Error & CNOT & Rot & Error & CNOT & Rot & Error & CNOT & Rot & Error & CNOT & Rot & Error & CNOT & Rot \\
\midrule
\textsc{CRLQAS} & 0.268 & 5 & 3 & 0.268 & 16 & 7 & 1.05 & 2 & 4 & 0.225 & 25 & 12 & $<$0.001 & 21 & 15 \\
\textsc{HyRLQAS} & 0.554 & 0 & 2 & 0.554 & 0 & 2 & 1.05 & 0 & 2 & 1.05 & 0 & 2 & $<$0.001 & 49 & 6 \\
\textsc{QuantumDARTS} & 0.279 & 50 & 12 & 0.366 & 56 & 6 & 0.282 & 39 & 12 & 1.05 & 2 & 6 & 5.66 & 13 & 6 \\
\textsc{GQEQAS} & 0.555 & 0 & 2 & 0.488 & 4 & 14 & 1.05 & 0 & 3 & 0.709 & 7 & 13 & $<$0.001 & 3 & 7 \\
\textsc{TFQAS} & 0.555 & 5 & 3 & 9.86 & 26 & 22 & 0.807 & 7 & 3 & 10.3 & 23 & 26 & 59.8 & 34 & 35 \\
\midrule
Method & \multicolumn{3}{c}{\texttt{H2\_Stretch}} & \multicolumn{3}{c}{\texttt{H2O\_StrongCorr}} & \multicolumn{3}{c}{\texttt{H4\_Chain}} & \multicolumn{3}{c}{\texttt{H3\_Linear}} & \multicolumn{3}{c}{\texttt{H3\_Linear}} \\
 & \multicolumn{3}{c}{} & \multicolumn{3}{c}{} & \multicolumn{3}{c}{all-to-all} & \multicolumn{3}{c}{all-to-all} & \multicolumn{3}{c}{linear} \\
\cmidrule(lr){2-4} \cmidrule(lr){5-7} \cmidrule(lr){8-10} \cmidrule(lr){11-13} \cmidrule(lr){14-16}
 & Error & CNOT & Rot & Error & CNOT & Rot & Error & CNOT & Rot & Error & CNOT & Rot & Error & CNOT & Rot \\
\midrule
\textsc{CRLQAS} & $<$0.001 & 4 & 5 & 6.46 & 42 & 28 & 62.0 & 45 & 25 & 9.25 & 36 & 34 & 23.5 & 24 & 46 \\
\textsc{HyRLQAS} & $<$0.001 & 14 & 7 & 46.7 & 33 & 37 & 67.8 & 63 & 7 & 33.5 & 33 & 17 & 46.4 & 25 & 25 \\
\textsc{QuantumDARTS} & $<$0.001 & 5 & 6 & 46.7 & 14 & 9 & 299 & 7 & 9 & 46.4 & 40 & 12 & 46.4 & 3 & 9 \\
\textsc{GQEQAS} & $<$0.001 & 3 & 9 & 46.7 & 1 & 4 & 67.8 & 2 & 7 & 33.5 & 2 & 14 & 46.4 & 1 & 14 \\
\textsc{TFQAS} & 4.95 & 26 & 24 & 269 & 36 & 29 & 201 & 33 & 37 & 307 & 27 & 23 & 72.0 & 21 & 25 \\
\bottomrule
\end{tabular}
\end{table*}

\begin{table*}[!tp]
\centering
\scriptsize
\caption{Best evaluation-stage circuit of every method on the regime R5 \texttt{BeH2} basis ladder: energy error (mHa), CNOT count, and rotation-gate count. The 6-qubit rung is the shallow-budget \texttt{BeH2\_STO3G} instance of Table~\ref{tab:eval_results}.}
\label{tab:eval_results_t5}
\setlength{\tabcolsep}{4pt}
\begin{tabular}{@{}lrrrrrrrrrrrrrrr@{}}
\toprule
Method & \multicolumn{3}{c}{\texttt{BeH2\_STO3G}} & \multicolumn{3}{c}{\texttt{BeH2\_631G}} & \multicolumn{3}{c}{\texttt{BeH2\_6311G}} & \multicolumn{3}{c}{\texttt{BeH2\_CCPVDZ\_12q}} & \multicolumn{3}{c}{\texttt{BeH2\_CCPVDZ\_14q}} \\
 & \multicolumn{3}{c}{6 qubits} & \multicolumn{3}{c}{8 qubits} & \multicolumn{3}{c}{10 qubits} & \multicolumn{3}{c}{12 qubits} & \multicolumn{3}{c}{14 qubits} \\
\cmidrule(lr){2-4} \cmidrule(lr){5-7} \cmidrule(lr){8-10} \cmidrule(lr){11-13} \cmidrule(lr){14-16}
 & Error & CNOT & Rot & Error & CNOT & Rot & Error & CNOT & Rot & Error & CNOT & Rot & Error & CNOT & Rot \\
\midrule
\textsc{CRLQAS} & 0.268 & 5 & 3 & 0.234 & 11 & 26 & 0.897 & 16 & 4 & 2.00 & 90 & 10 & 6.15 & 75 & 25 \\
\textsc{HyRLQAS} & 0.554 & 0 & 2 & 2.17 & 54 & 16 & 0.897 & 2 & 1 & 2.00 & 96 & 4 & 76.3 & 82 & 18 \\
\textsc{QuantumDARTS} & 0.279 & 50 & 12 & 2.17 & 5 & 3 & 0.897 & 0 & 9 & 2.00 & 167 & 27 & 274 & 231 & 18 \\
\textsc{GQEQAS} & 0.555 & 0 & 2 & 2.17 & 0 & 2 & 0.897 & 0 & 2 & 1.83 & 5 & 7 & 6.13 & 5 & 6 \\
\textsc{TFQAS} & 0.555 & 5 & 3 & 135 & 38 & 29 & 229 & 45 & 43 & 231 & 57 & 31 & 6.17 & 40 & 19 \\
\bottomrule
\end{tabular}
\end{table*}

\begin{table}[H]
\centering
\footnotesize
\caption{Gate redundancy $\rho$ of the chemically accurate evaluation-stage circuits on \texttt{BeH2\_STO3G}, one circuit per seed, computed from the mean gate count against the two-gate reference (Table~\ref{tab:tier1_full}). --- indicates that no evaluation-stage circuit reached chemical accuracy under that budget.}
\label{tab:tier1_redundancy}
\setlength{\tabcolsep}{8pt}
\begin{tabular}{@{}lrr@{}}
\toprule
Method & Shallow ($D=10$) & Deep ($D=50$) \\
\midrule
\textsc{HyRLQAS} & 0.0\% & 0.0\% \\
\textsc{GQEQAS} & 0.0\% & 83.3\% \\
\textsc{CRLQAS} & 75.0\% & 88.0\% \\
\textsc{TFQAS} & 76.9\% & --- \\
\textsc{QuantumDARTS} & 93.2\% & 95.0\% \\
\bottomrule
\end{tabular}
\end{table}

\begin{table*}[!tp]
\centering
\footnotesize
\caption{Per-qubit entropy $S(q)$ (bits) on \texttt{H2O\_StrongCorr} for the best evaluation-stage circuit of every method, against the exact ground state. Bold \textbf{.000}: $S(q) < 10^{-6}$\,bit. $\varepsilon$ in mHa.}
\label{tab:h2o_perqubit}
\setlength{\tabcolsep}{5pt}
\begin{tabular}{@{}lrrrrrrrrrr@{}}
\toprule
Source & $\varepsilon$ & $q_0$ & $q_1$ & $q_2$ & $q_3$ & $q_4$ & $q_5$ & $q_6$ & $q_7$ & $\MAES$ \\
\midrule
Exact & --- & .021 & .021 & .342 & .342 & .316 & .316 & .071 & .071 & --- \\
\textsc{CRLQAS} & 6.5 & \textbf{.000} & \textbf{.000} & .331 & .331 & .331 & .331 & \textbf{.000} & \textbf{.000} & 0.030 \\
\textsc{HyRLQAS} & 46.7 & \textbf{.000} & \textbf{.000} & \textbf{.000} & \textbf{.000} & \textbf{.000} & \textbf{.000} & \textbf{.000} & \textbf{.000} & 0.187 \\
\textsc{QuantumDARTS} & 46.7 & \textbf{.000} & \textbf{.000} & \textbf{.000} & \textbf{.000} & \textbf{.000} & \textbf{.000} & \textbf{.000} & \textbf{.000} & 0.187 \\
\textsc{GQEQAS} & 46.7 & \textbf{.000} & \textbf{.000} & \textbf{.000} & \textbf{.000} & \textbf{.000} & \textbf{.000} & \textbf{.000} & \textbf{.000} & 0.187 \\
\textsc{TFQAS} & 269.1 & .073 & .104 & .063 & .685 & .102 & .685 & .681 & .057 & 0.246 \\
\bottomrule
\end{tabular}
\end{table*}

\begin{table*}[!tp]
\centering
\footnotesize
\caption{Per-qubit von Neumann entropy $S(q)$ (bits) on \texttt{H4\_Chain} for the best evaluation-stage circuit of every method under all-to-all and nearest-neighbour (linear) connectivity, against the exact ground state. Bold \textbf{.000}: $S(q) < 10^{-6}$\,bit. Provenance as in Table~\ref{tab:entropy_consistency}. Rows that are bold on every qubit are product-state outputs (Section~\ref{sec:local_entropy}); their $\MAES$ equals $\bar{S}_{\mathrm{exact}} = 0.198$.}
\label{tab:h4_perqubit_all}
\setlength{\tabcolsep}{4pt}
\begin{tabular}{@{}llrrrrrrrrrr@{}}
\toprule
Source & Connectivity & $\varepsilon$ (mHa) & $q_0$ & $q_1$ & $q_2$ & $q_3$ & $q_4$ & $q_5$ & $q_6$ & $q_7$ & $\MAES$ \\
\midrule
Exact & --- & --- & .124 & .124 & .274 & .274 & .284 & .284 & .109 & .109 & --- \\
\midrule
\textsc{CRLQAS} & all-to-all & 62.0 & \textbf{.000} & \textbf{.000} & .030 & .030 & \textbf{.000} & \textbf{.000} & .030 & .030 & 0.183 \\
\textsc{HyRLQAS} & all-to-all & 67.8 & \textbf{.000} & \textbf{.000} & \textbf{.000} & \textbf{.000} & \textbf{.000} & \textbf{.000} & \textbf{.000} & \textbf{.000} & 0.198 \\
\textsc{QuantumDARTS} & all-to-all & 299.2 & \textbf{.000} & \textbf{.000} & \textbf{.000} & \textbf{.000} & \textbf{.000} & \textbf{.000} & \textbf{.000} & \textbf{.000} & 0.198 \\
\textsc{GQEQAS} & all-to-all & 67.8 & \textbf{.000} & \textbf{.000} & \textbf{.000} & \textbf{.000} & \textbf{.000} & \textbf{.000} & \textbf{.000} & \textbf{.000} & 0.198 \\
\textsc{TFQAS} & all-to-all & 200.7 & .047 & .017 & .019 & .119 & .096 & .095 & .001 & .001 & 0.148 \\
\midrule
\textsc{CRLQAS} & linear & 67.8 & \textbf{.000} & \textbf{.000} & \textbf{.000} & \textbf{.000} & \textbf{.000} & \textbf{.000} & \textbf{.000} & \textbf{.000} & 0.198 \\
\textsc{HyRLQAS} & linear & 67.8 & \textbf{.000} & \textbf{.000} & \textbf{.000} & \textbf{.000} & \textbf{.000} & \textbf{.000} & \textbf{.000} & \textbf{.000} & 0.198 \\
\textsc{QuantumDARTS} & linear & 67.8 & \textbf{.000} & \textbf{.000} & \textbf{.000} & \textbf{.000} & \textbf{.000} & \textbf{.000} & \textbf{.000} & \textbf{.000} & 0.198 \\
\textsc{GQEQAS} & linear & 67.8 & \textbf{.000} & \textbf{.000} & \textbf{.000} & \textbf{.000} & \textbf{.000} & \textbf{.000} & \textbf{.000} & \textbf{.000} & 0.198 \\
\textsc{TFQAS} & linear & 365.5 & .167 & .412 & .468 & .054 & .147 & .220 & \textbf{.000} & \textbf{.000} & 0.146 \\
\bottomrule
\end{tabular}
\end{table*}

\subsection{Training and Evaluation Outputs}
\label{App:training_results}

Tables~\ref{tab:main_results} and~\ref{tab:main_results_t5} report the best training-stage circuits of the two reinforcement-learning methods. For each configuration, we take the lowest recorded error across all seeds and training snapshots, together with the CNOT and rotation counts of the circuit that attained it. Reconstruction reproduces the recorded energies. The other three methods export one circuit per seed and are omitted from these training-stage tables.

Comparison with the evaluation-stage outputs in Tables~\ref{tab:eval_results} and~\ref{tab:eval_results_t5} shows two patterns.

On Regime~R1 and on \texttt{H2\_Stretch}, both stages reach chemical accuracy for both methods.

On harder instances, exploration often finds lower-error circuits than the selected checkpoints return. For example, \textsc{CRLQAS} reaches $2.86$\,mHa on \texttt{H2O\_StrongCorr} during training, compared with $6.46$\,mHa at evaluation. Both reinforcement-learning methods show this pattern on \texttt{H4\_Chain}, \texttt{H3\_Linear} with all-to-all connectivity, and the 12- and 14-qubit \texttt{BeH2} instances. Under linear connectivity on \texttt{H3\_Linear}, the two stages coincide for \textsc{CRLQAS} at $23.5$\,mHa.

\begin{table*}[!tp]
\centering
\scriptsize
\caption{Best training-stage circuit of the two reinforcement-learning methods on regimes R1--R4: energy error (mHa), CNOT count, and rotation-gate count. Errors below $0.001$\,mHa are shown as $<$0.001.}
\label{tab:main_results}
\setlength{\tabcolsep}{4pt}
\begin{tabular}{@{}lrrrrrrrrrrrrrrr@{}}
\toprule
Method & \multicolumn{3}{c}{\texttt{BeH2\_STO3G}} & \multicolumn{3}{c}{\texttt{BeH2\_STO3G}} & \multicolumn{3}{c}{\texttt{LiH\_Equil}} & \multicolumn{3}{c}{\texttt{LiH\_Equil}} & \multicolumn{3}{c}{\texttt{CH2}} \\
 & \multicolumn{3}{c}{$D=10$} & \multicolumn{3}{c}{$D=50$} & \multicolumn{3}{c}{$D=10$} & \multicolumn{3}{c}{$D=50$} & \multicolumn{3}{c}{} \\
\cmidrule(lr){2-4} \cmidrule(lr){5-7} \cmidrule(lr){8-10} \cmidrule(lr){11-13} \cmidrule(lr){14-16}
 & Error & CNOT & Rot & Error & CNOT & Rot & Error & CNOT & Rot & Error & CNOT & Rot & Error & CNOT & Rot \\
\midrule
\textsc{CRLQAS} & 0.554 & 4 & 3 & 0.269 & 30 & 10 & 0.948 & 4 & 6 & 0.229 & 13 & 8 & $<$0.001 & 35 & 23 \\
\textsc{HyRLQAS} & 0.484 & 6 & 3 & 0.298 & 13 & 8 & 1.05 & 3 & 4 & 0.235 & 11 & 7 & $<$0.001 & 9 & 49 \\
\midrule
Method & \multicolumn{3}{c}{\texttt{H2\_Stretch}} & \multicolumn{3}{c}{\texttt{H2O\_StrongCorr}} & \multicolumn{3}{c}{\texttt{H4\_Chain}} & \multicolumn{3}{c}{\texttt{H3\_Linear}} & \multicolumn{3}{c}{\texttt{H3\_Linear}} \\
 & \multicolumn{3}{c}{} & \multicolumn{3}{c}{} & \multicolumn{3}{c}{all-to-all} & \multicolumn{3}{c}{all-to-all} & \multicolumn{3}{c}{linear} \\
\cmidrule(lr){2-4} \cmidrule(lr){5-7} \cmidrule(lr){8-10} \cmidrule(lr){11-13} \cmidrule(lr){14-16}
 & Error & CNOT & Rot & Error & CNOT & Rot & Error & CNOT & Rot & Error & CNOT & Rot & Error & CNOT & Rot \\
\midrule
\textsc{CRLQAS} & $<$0.001 & 10 & 6 & 2.86 & 33 & 21 & 33.5 & 47 & 10 & 3.42 & 36 & 25 & 23.5 & 8 & 37 \\
\textsc{HyRLQAS} & $<$0.001 & 6 & 7 & 6.76 & 39 & 13 & 33.7 & 36 & 12 & 5.57 & 18 & 13 & 23.5 & 8 & 6 \\
\bottomrule
\end{tabular}
\end{table*}

\begin{table*}[!tp]
\centering
\scriptsize
\caption{Best training-stage circuit of the two reinforcement-learning methods on the regime R5 \texttt{BeH2} ladder: energy error (mHa), CNOT count, and rotation-gate count.}
\label{tab:main_results_t5}
\setlength{\tabcolsep}{4pt}
\begin{tabular}{@{}lrrrrrrrrrrrrrrr@{}}
\toprule
Method & \multicolumn{3}{c}{\texttt{BeH2\_STO3G}} & \multicolumn{3}{c}{\texttt{BeH2\_631G}} & \multicolumn{3}{c}{\texttt{BeH2\_6311G}} & \multicolumn{3}{c}{\texttt{BeH2\_CCPVDZ\_12q}} & \multicolumn{3}{c}{\texttt{BeH2\_CCPVDZ\_14q}} \\
 & \multicolumn{3}{c}{6 qubits} & \multicolumn{3}{c}{8 qubits} & \multicolumn{3}{c}{10 qubits} & \multicolumn{3}{c}{12 qubits} & \multicolumn{3}{c}{14 qubits} \\
\cmidrule(lr){2-4} \cmidrule(lr){5-7} \cmidrule(lr){8-10} \cmidrule(lr){11-13} \cmidrule(lr){14-16}
 & Error & CNOT & Rot & Error & CNOT & Rot & Error & CNOT & Rot & Error & CNOT & Rot & Error & CNOT & Rot \\
\midrule
\textsc{CRLQAS} & 0.554 & 4 & 3 & 0.115 & 38 & 10 & 0.126 & 58 & 25 & 1.24 & 69 & 19 & 4.99 & 77 & 14 \\
\textsc{HyRLQAS} & 0.484 & 6 & 3 & 0.234 & 16 & 9 & 0.897 & 38 & 8 & 1.24 & 24 & 4 & 5.00 & 80 & 17 \\
\bottomrule
\end{tabular}
\end{table*}

\subsection{Regime~R1 Full Gate Statistics}
\label{App:tier1_full}

Table~\ref{tab:tier1_full} gives gate statistics for chemically accurate evaluation-stage circuits on both Regime~R1 instances and budgets. Each seed contributes one circuit: the reconstructed rollout for reinforcement-learning methods or the exported circuit otherwise. Redundancy is calculated from the mean gate count relative to the two-gate reference.

\textsc{HyRLQAS} returns two gates in every seed and setting. \textsc{GQEQAS} does so on \texttt{BeH2\_STO3G} with the shallow budget; its deep-budget means are 12.0 gates on \texttt{BeH2\_STO3G} and 20.5 on \texttt{LiH\_Equil}. \textsc{QuantumDARTS} exceeds 83\% redundancy in every setting. Its widest range across seeds is 13--62 gates, on shallow-budget \texttt{BeH2\_STO3G}.

\label{App:tier1_gate_stats}

\begin{table*}[!tp]
\centering
\footnotesize
\caption{Gate statistics of the chemically accurate evaluation-stage circuits in Regime R1, one circuit per seed: number of such circuits $N$, mean rotation-gate count (Rot), mean CNOT count, mean and range of the total gate count, and the redundancy of the mean gate count against the two-gate reference. --- indicates that no evaluation-stage circuit reached chemical accuracy under that budget.}
\label{tab:tier1_full}
\setlength{\tabcolsep}{6pt}
\begin{tabular}{@{}llrrrrrr@{}}
\toprule
Method & Budget & $N$ & Rot mean & CNOT mean & Total mean & Total range & Redundancy \\
\midrule
\multicolumn{8}{l}{\textit{BeH$_2$ STO-3G}} \\
\addlinespace[1pt]
\textsc{CRLQAS} & shallow & 3 & 3.00 & 5.00 & 8.00 & 8--8 & 75.0\% \\
\textsc{HyRLQAS} & shallow & 3 & 2.00 & 0.00 & 2.00 & 2--2 & 0.0\% \\
\textsc{GQEQAS} & shallow & 2 & 2.00 & 0.00 & 2.00 & 2--2 & 0.0\% \\
\textsc{TFQAS} & shallow & 3 & 3.33 & 5.33 & 8.67 & 8--9 & 76.9\% \\
\textsc{QuantumDARTS} & shallow & 3 & 7.00 & 22.33 & 29.33 & 13--62 & 93.2\% \\
\addlinespace[2pt]
\textsc{CRLQAS} & deep & 3 & 6.00 & 10.67 & 16.67 & 13--23 & 88.0\% \\
\textsc{HyRLQAS} & deep & 3 & 2.00 & 0.00 & 2.00 & 2--2 & 0.0\% \\
\textsc{GQEQAS} & deep & 2 & 10.00 & 2.00 & 12.00 & 6--18 & 83.3\% \\
\textsc{TFQAS} & deep & 0 & --- & --- & --- & --- & --- \\
\textsc{QuantumDARTS} & deep & 3 & 9.00 & 30.67 & 39.67 & 27--62 & 95.0\% \\
\midrule
\multicolumn{8}{l}{\textit{LiH (equil.)}} \\
\addlinespace[1pt]
\textsc{CRLQAS} & shallow & 3 & 4.33 & 2.00 & 6.33 & 6--7 & 68.4\% \\
\textsc{HyRLQAS} & shallow & 3 & 2.00 & 0.00 & 2.00 & 2--2 & 0.0\% \\
\textsc{GQEQAS} & shallow & 2 & 4.00 & 0.00 & 4.00 & 3--5 & 50.0\% \\
\textsc{TFQAS} & shallow & 3 & 4.00 & 5.67 & 9.67 & 9--10 & 79.3\% \\
\textsc{QuantumDARTS} & shallow & 3 & 7.00 & 18.33 & 25.33 & 12--51 & 92.1\% \\
\addlinespace[2pt]
\textsc{CRLQAS} & deep & 3 & 10.67 & 20.67 & 31.33 & 25--37 & 93.6\% \\
\textsc{HyRLQAS} & deep & 3 & 2.00 & 0.00 & 2.00 & 2--2 & 0.0\% \\
\textsc{GQEQAS} & deep & 2 & 14.00 & 6.50 & 20.50 & 20--21 & 90.2\% \\
\textsc{TFQAS} & deep & 0 & --- & --- & --- & --- & --- \\
\textsc{QuantumDARTS} & deep & 2 & 7.50 & 4.50 & 12.00 & 8--16 & 83.3\% \\
\bottomrule
\end{tabular}
\end{table*}

\subsection{Topology Analysis: \texttt{H\textsubscript{4} Chain} Connectivity 
Comparison}
\label{App:h4_topology}

The connectivity effect on the entropy profile is measured on \texttt{H4\_Chain}, whose ground state is unique, under two connectivity settings: all-to-all and nearest-neighbour linear.

Table~\ref{tab:h4_stages} identifies which circuit stage each connectivity table measures. For pruned training-stage circuits, linear connectivity raises $\MAES$ from 0.068 to 0.121 and the energy error from 33.55 to 48.03\,mHa (Table~\ref{tab:h4_connectivity}). Before pruning, the corresponding \textsc{CRLQAS} values are 0.068 and 0.102, with errors of 33.5 and 43.7\,mHa (Table~\ref{tab:h4_perqubit_train}). Pruning removes gates and re-optimizes parameters, accounting for these differences.

\begin{table*}[!tp]
\centering
\begin{minipage}[t]{0.48\textwidth}
\centering
\footnotesize
\caption{Entropy-profile consistency for \texttt{H4\_Chain} under two connectivity settings, \textsc{CRLQAS} pruned training-stage circuits. $\mathrm{MAE}_S$ measures deviation from the exact ground-state per-qubit entropy profile.}
\label{tab:h4_connectivity}
\setlength{\tabcolsep}{6pt}
\begin{tabular}{@{}lrrrr@{}}
\toprule
Connectivity & Error (mHa) & $\bar{S}_{\mathrm{exact}}$ & $\bar{S}_{\mathrm{circuit}}$ & $\mathrm{MAE}_S$ \\
\midrule
All-to-all & 33.55 & 0.198 & 0.130 & 0.068 \\
Linear     & 48.03 & 0.198 & 0.077 & 0.121 \\
\bottomrule
\end{tabular}
\end{minipage}
\hfill
\begin{minipage}[t]{0.48\textwidth}
\centering
\footnotesize
\caption{The three \texttt{H4\_Chain} connectivity tables and the circuit stage each measures.}
\label{tab:h4_stages}
\setlength{\tabcolsep}{3pt}
\begin{tabular}{@{}lll@{}}
\toprule
Table & Circuit stage & Methods \\
\midrule
Table~\ref{tab:h4_perqubit_all} & evaluation stage (Section~\ref{sec:reconstruction}) & all five \\
Table~\ref{tab:h4_perqubit_train} & training-stage best (Section~\ref{sec:protocol_inputs}) & \textsc{CRLQAS}, \textsc{HyRLQAS} \\
Table~\ref{tab:h4_connectivity} & pruned training-stage & \textsc{CRLQAS} \\
\bottomrule
\end{tabular}
\end{minipage}
\end{table*}

\begin{table*}[!tp]
\centering
\footnotesize
\caption{As Table~\ref{tab:h4_perqubit_all}, for the \emph{training-stage best} circuits of the two reinforcement-learning methods (reported separately; see Section~\ref{sec:protocol_inputs}).}
\label{tab:h4_perqubit_train}
\setlength{\tabcolsep}{4pt}
\begin{tabular}{@{}llrrrrrrrrrr@{}}
\toprule
Source & Connectivity & $\varepsilon$ (mHa) & $q_0$ & $q_1$ & $q_2$ & $q_3$ & $q_4$ & $q_5$ & $q_6$ & $q_7$ & $\MAES$ \\
\midrule
Exact & --- & --- & .124 & .124 & .274 & .274 & .284 & .284 & .109 & .109 & --- \\
\midrule
\textsc{CRLQAS} & all-to-all & 33.5 & .062 & \textbf{.000} & .213 & .244 & .244 & .213 & \textbf{.000} & .062 & 0.068 \\
\textsc{HyRLQAS} & all-to-all & 33.7 & \textbf{.000} & .074 & .244 & .204 & .204 & .244 & .074 & \textbf{.000} & 0.067 \\
\textsc{CRLQAS} & linear & 43.7 & .002 & \textbf{.000} & .191 & .190 & .191 & .190 & .002 & \textbf{.000} & 0.102 \\
\textsc{HyRLQAS} & linear & 43.9 & \textbf{.000} & \textbf{.000} & .192 & .192 & .192 & .192 & \textbf{.000} & \textbf{.000} & 0.102 \\
\bottomrule
\end{tabular}
\end{table*}

Training-stage circuits retain entropy on the interior qubits $q_2$--$q_5$ under both connectivities (Table~\ref{tab:h4_perqubit_train}). With all-to-all connectivity, two boundary qubits fall below the entropy threshold. With linear connectivity, interior entropies are about $0.19$\,bit and all four boundary entropies are at most $0.002$\,bit. For \textsc{CRLQAS}, the corresponding energy error increases from 33.5 to 43.7\,mHa.

Evaluation-stage outputs differ from these training snapshots (Table~\ref{tab:h4_perqubit_all}). Both methods return product states under linear connectivity, and \textsc{HyRLQAS} also does so under all-to-all connectivity. These circuits contain 12--63 CNOT gates. Thus the entangled states found during training are not retained in these evaluation outputs.

\begin{table*}[!tp]
\centering
\footnotesize
\caption{Evaluation-stage results on the \texttt{BeH2} basis ladder (Regime R5). Best: lowest energy error of the evaluation-stage circuit (mHa; after re-optimization for \textsc{CRLQAS} and \textsc{HyRLQAS}, Section~\ref{sec:reconstruction}). SR: success rate with its numerator and denominator. For \textsc{CRLQAS} and \textsc{HyRLQAS} it is the rollout success rate, the recorded fraction of rollouts at chemical accuracy at the selected checkpoint, pooled over the evaluation runs of that instance (20 rollouts each); for the other methods it is the run success rate, the fraction of seeds whose evaluation-stage circuit is at chemical accuracy. The two rates have different units.}
\label{tab:tier5_full}
\setlength{\tabcolsep}{6pt}
\renewcommand{\arraystretch}{1.12}
\begin{tabular}{@{}lrlrlrlrl@{}}
\toprule
& \multicolumn{2}{c}{8q} & \multicolumn{2}{c}{10q} & \multicolumn{2}{c}{12q} & \multicolumn{2}{c}{14q} \\
\cmidrule(lr){2-3}\cmidrule(lr){4-5}\cmidrule(lr){6-7}\cmidrule(lr){8-9}
Method & Best & SR & Best & SR & Best & SR & Best & SR \\
\midrule
\textsc{CRLQAS} & 0.23 & 1/40 (2.5\%) & 0.90 & 31/40 (78\%) & 2.00 & 0/20 & 6.15 & 0/60 \\
\textsc{HyRLQAS} & 2.17 & 0/60 & 0.90 & 40/40 (100\%) & 2.00 & 0/40 & 76.3 & 0/60 \\
\textsc{QuantumDARTS} & 2.17 & 0/3 & 0.90 & 3/3 (100\%) & 2.00 & 0/3 & 273.8 & 0/3 \\
\textsc{GQEQAS} & 2.17 & 0/2 & 0.90 & 2/2 (100\%) & 1.83 & 0/2 & 6.13 & 0/2 \\
\textsc{TFQAS} & 134.7 & 0/3 & 229.4 & 0/3 & 231.1 & 0/3 & 6.17 & 0/3 \\
\bottomrule
\end{tabular}
\end{table*}

\subsection{Regime~R5 Full Scalability Results}
\label{App:tier5_full}

Table~\ref{tab:tier5_full} pairs each method's evaluation-stage success rate with its best error on the \texttt{BeH2} ladder. Table~\ref{tab:main_results_t5} separately reports the best training-stage circuits. The following comparisons supplement Section~\ref{sec:r5_results}.

\paragraph{8 qubits.} The best reconstructed \textsc{CRLQAS} evaluation circuit reaches $0.23$\,mHa. At the selected checkpoints, 1 of 40 recorded rollouts reaches chemical accuracy. No other method reaches the threshold; three return product-state outputs at $2.17$\,mHa.

\paragraph{12 qubits.} \textsc{CRLQAS} reaches $1.24$\,mHa during training, but its evaluation output has the product-state error of $2.00$\,mHa. The extraction analysis finds six such training circuits, with 8--64 gates and pruning ratios of $25$--$78\%$. The selected evaluation checkpoint does not return these chemically accurate circuits.

\paragraph{14 qubits.} The best training-stage errors, $4.99$ and $5.00$\,mHa, remain outside chemical accuracy, although they are lower than the evaluation-stage product-state errors.

\clearpage                         

\section*{Data and Code Availability}
The repository at \url{https://github.com/jiayangniu/HamQASBench} contains the benchmark instances, exact reference states, and generation scripts. It also provides the method implementations, execution backend, pruning procedure, and entropy-analysis tools. The released per-run data support every table and figure and include the complete reporting table specified in Section~\ref{sec:reporting}. An archival snapshot will be deposited at publication.

\bibliographystyle{IEEEtran}
\bibliography{references}

% ...

\EOD

\end{document}